\pdfoutput=1
\documentclass[prx,twocolumn,notitlepage,superscriptaddress,amsmath,amssymb,aps,floatfix,longbibliography]{revtex4-1}
\usepackage[T1]{fontenc}
\usepackage{color}
\usepackage{amsmath}
\usepackage{amssymb}
\usepackage{bm}
\usepackage{graphicx}
\usepackage[usenames,dvipsnames]{xcolor}
\usepackage[%
  colorlinks=true,
  urlcolor=blue,
  linkcolor=blue,
  citecolor=blue
]{hyperref}
\usepackage[normalem]{ulem}
\usepackage{array}
\newcommand{\fupdate}[1]{\textcolor{black}{#1}}

\newcommand{\EDF}{Appendix Fig.}% to use for submission to arxiv

\begin{document}
%TC:ignore
\pdfoutput=1
\title{Programming nanomechanical computation with light}
\author{Xiaofei Guo}
\affiliation{Department of Information in Matter, AMOLF, Science Park 104, 1098 XG Amsterdam, The Netherlands}
\author{Jonne Drost}
\affiliation{Department of Information in Matter, AMOLF, Science Park 104, 1098 XG Amsterdam, The Netherlands}
\author{Fons van der Laan}
\affiliation{Department of Information in Matter, AMOLF, Science Park 104, 1098 XG Amsterdam, The Netherlands}
\affiliation{QSTeM, Testbed for Mechanical Quantum Sensing, Delftechpark 01, 2628 XJ Delft, The Netherlands}
\author{Jesse J. Slim}
\affiliation{Australian Research Council Centre of Excellence for Engineered Quantum Systems (EQUS), School of Mathematics and Physics, University of Queensland, St Lucia, QLD, Australia}
\affiliation{Department of Information in Matter, AMOLF, Science Park 104, 1098 XG Amsterdam, The Netherlands}
\author{Marc Serra-Garcia}
\email{M.SerraGarcia@amolf.nl}
\affiliation{Department of Information in Matter, AMOLF, Science Park 104, 1098 XG Amsterdam, The Netherlands}
\author{Ewold Verhagen}
\email{E.Verhagen@amolf.nl}
\affiliation{Department of Information in Matter, AMOLF, Science Park 104, 1098 XG Amsterdam, The Netherlands}

\begin{abstract} 
Looking at physical systems as computers allows us to regard physical properties, such as thermal noise, symmetry or topology, as unconventional resources for computation. However, harnessing these resources requires programming computational functionality through strong, controllable nonlinearities in the system. Here, we show that cavity optomechanical interactions allow laser-controlled computation with nanomechanical degrees of freedom. We demonstrate a set of basic digital logic gates with level restoration and controlled mechanical couplings as essential ingredients for arbitrary computing networks. Owing to the strong optomechanical nonlinearity and precise readout, the system operates close to thermal amplitudes, in the regime where thermodynamic stochasticity governs its behavior. This opens a new path for the realization of physical computing with controlled nonlinear resonators.
\end{abstract}

\maketitle

Physical computing, understood as the use of physical dynamics and interactions to perform information processing, provides novel viewpoints on the nature, limits, and possibilities of information processing. It is pursued for e.g. specific edge computing applications, neuromorphic or probabilistic computing paradigms, and fundamental studies of computing performance in minimal systems~\cite{roukes2004mechanical,vandoorne2014experimental, yasuda2021mechanical,kaspar2021rise,wright2022deep,mcmahon2023physics,  dubvcek2024sensor,romero2024acoustically,zhou2025harnessing,Camsari2019p-bits,lopez2016sub, landauer1961irreversibility, berut2012experimental}. Computing functionalities can be recognized or engineered in a vast range of physical systems beyond electronics, from soft matter and biology to photonics and mechanics~\cite{roukes2004mechanical,badzey2004controllable,prakash2007microfluidic,guerra2010noise,mahboob2011interconnect,moon2012genetic,vandoorne2014experimental,fredkin2014,yao2014logic,mahboob2014multimode,lopez2016sub,raney2016stable,dion2018reservoir,ilyas2019cascadable,song2019additively,el2021digital,mei2021mechanical,yasuda2021mechanical,lv2023dna,mcmahon2023physics, he2024programmable, wang2024harnessing,byun2024integrated,dubvcek2024sensor,romero2024acoustically,lee2024task,zhou2025harnessing,song2025heat, watkins2025arbitrary}. Specifically, networks of resonators offer well-defined sites and interactions in which physical principles such as symmetry, topology, and fluctuations can be harnessed for computation~\cite{schmid2016fundamentals, roukes2004mechanical, ilyas2019cascadable, guerra2010noise, badzey2004controllable, fredkin2014, mahboob2011interconnect, yao2014logic, mahboob2014multimode, coulombe2017computing, tadokoro2021highly,eichler2012strong,rieser2022tunable,jin2024engineering,venkatesh2017implementation, deshaka2025realization}.

Central to practically all computing functionality is nonlinearity: Strong, controlled nonlinearities are essential in paradigms ranging from recurrent neural networks and reservoir computers to digital processing, and offer possibilities such as level restoration~\cite{jin2024engineering} and reliable cascading of computational elements. 
In most physical computing networks, however, nonlinearities and network topologies are fixed by intrinsic material and geometric properties, necessitating device redesign and refabrication to change performance~\cite{song2019additively, raney2016stable, byun2024integrated, el2021digital,guerra2010noise, badzey2004controllable, yao2014logic, dion2018reservoir}. This can also limit scalability to larger computing networks, for example by requiring redefinition of digital logic states (binary 0 and 1)~\cite{guerra2010noise, fredkin2014, mahboob2011interconnect}. Ideal resonator-based physical computing platforms could instead be fully programmed in situ, both in nonlinear function and network connectivity --- in analogy to programmable electronics such as FPGAs.

%TC:ignore
\begin{figure*}[t]
    \centering
    \includegraphics[width=\textwidth]{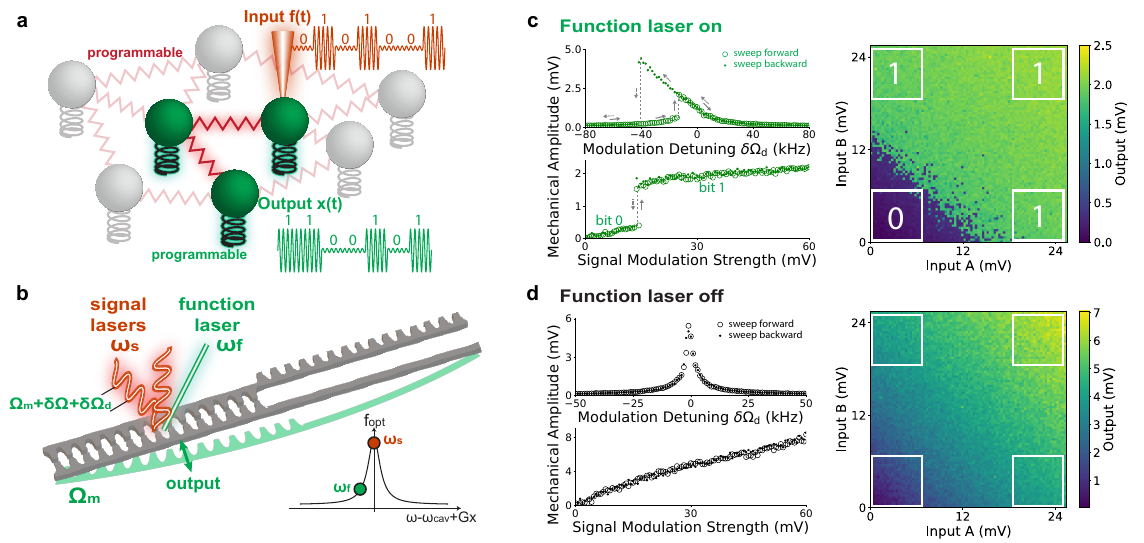}
    \caption{{\bf Computation and programmable nonlinearity enabled by optomechanical interactions.} {\bf a.} Schematic of computation using an oscillator network. A programmable response of individual oscillators and their interactions enables the network output (oscillator motion) to be tuned to a desired target under a prescribed external-force input. {\bf b.} Optomechanical system used to realize the oscillator network. A nanobeam (gray) serves as the mechanical oscillators, the second eigenmode (green) is read out as the output. The inputs are provided by signal lasers with intensity modulation at $\Omega_\mathrm{d} = \Omega_m+\delta\Omega+\delta\Omega_\mathrm{d}$, and a function laser programs the properties of individual oscillators. Here $\delta\Omega$ denotes the effective spring effect of the function laser, as described in Eq.~\ref{eq:springshift} and Sec.~\ref{sec:theory_singlemode}. {\bf c.} and {\bf d.} The function laser enables programmable switching between linear and nonlinear response. Left: Measured mechanical response while sweeping the modulation detuning and modulation strength of the signal laser. Right: With the function laser on, the induced nonlinearity enables an OR gate with inputs defined by the signal-laser modulation strengths; with the function laser off, the linear mechanical response cannot realize a logic gate.}
    \label{fig:Figure1}
\end{figure*}
%TC:endignore

%TC:ignore
\begin{figure*}[t]
    \centering
    \includegraphics[width=0.8\textwidth]{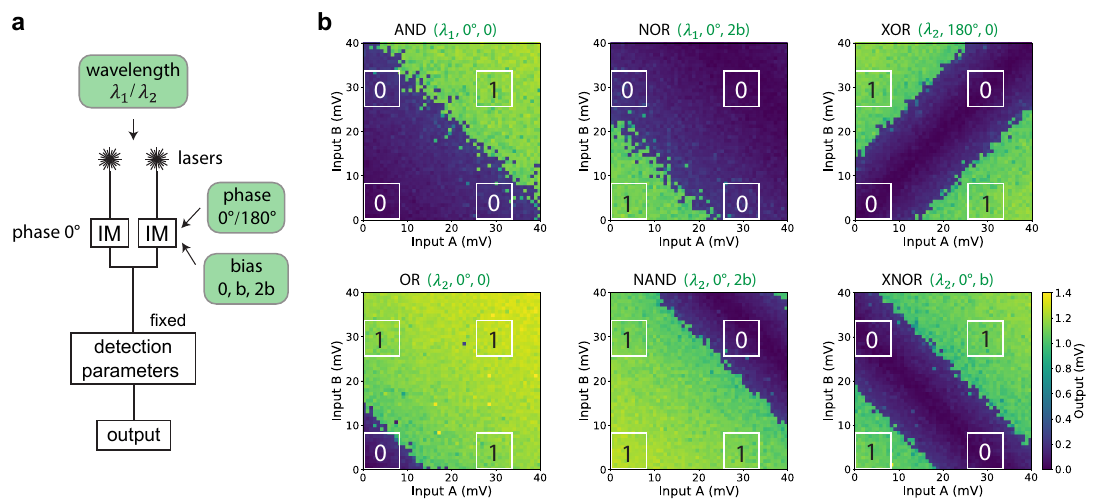}
    \caption{{\bf Reconfigurable implementation of six standard Boolean operations in a single device.} {\bf a.} The device function is reconfigured by three control parameters: the laser wavelength $\lambda$, the modulator phase, and the modulation amplitude bias. The modulation bias is applied with 180$^\circ$ phase. {\bf b.} 
    Without redefining the input and output convention, we experimentally realize six standard Boolean operations by appropriately tuning these three parameters.}
    \label{fig:Figure2}
\end{figure*}
%TC:endignore

In this work, we present an approach for physical computing that takes one or more linear, uncoupled nanomechanical resonators and endows them with computational functionality by programming nonlinearity and couplings through cavity optomechanical interactions. It exploits radiation pressure backaction to induce strong nanomechanical nonlinearity through suitable laser driving. This enables information processing on resonator vibrational states that can be reconfigured in real time without device redesign. We illustrate this approach by implementing digital logic based on Duffing nonlinearity. Using laser programming, we realize all six fundamental logic operations in a single degree of freedom, without redefining digital thresholds for bits 0 and 1. 
We further show that mechanical eigenmodes of different orders can be used as distinct degrees of freedom and coupled through individually controllable optical interactions~\cite{mathew2020synthetic, delpino2022, Bagheri2013Photonic}, enabling real-time programmable nonlinear networks. We demonstrate this capability in a three-mode computation task. 
Importantly, the optically induced nonlinearity in our platform exceeds the (fixed) intrinsic nonlinearity by orders of magnitude, allowing digital logic at sub-nanometer amplitudes, only a few times the amplitude of thermal fluctuations. Combined with the sensitive readout enabled by optomechanics, we demonstrate that the gate fidelity is governed by thermal fluctuations from the room-temperature environment, and probe the fundamental trade-off between computational fidelity and operation speed.

\subsection*{Computation through programmable optomechanical nonlinearity}

In a mechanical resonator network performing a computational task, as envisioned in Fig.~\ref{fig:Figure1}a, an external signal $f(t)$, resonant with one or more nodes, serves as input information carrier. Through the collective linear and/or nonlinear dynamics of coupled resonators, the network transforms this force input into a target oscillatory displacement output $x(t)$.
To make a single network useful for a wide range of tasks, one seeks to program its full response, defined through the nonlinear network potential. This includes reconfiguring the network's coupling configuration, i.e., the interaction strengths between nodes, and tuning the nonlinear dynamical response of each oscillator.

We address the challenge of programmability using the mechanism of cavity optomechanical backaction~\cite{aspelmeyer2014cavity}, which enables real-time laser control of both dynamics and couplings of mechanical resonators. The mechanisms described below could in principle be implemented in any optomechanical resonator in the unresolved-sideband regime (with mechanical frequencies smaller than the cavity linewidth $\kappa$). We implement them in a pair of silicon nanobeams, shown in Fig.~\ref{fig:Figure1}b, whose flexural eigenmodes serve as optically controllable mechanical degrees of freedom. These modes are parametrically coupled to the optical field of a sliced photonic-crystal nanocavity confined between the teeth of both beams, such that a local mode displacement $x$ from equilibrium shifts the cavity resonance frequency by $Gx$, with $G$ the optomechanical coupling strength. 

We now discuss how a laser can imprint strong mechanical nonlinearity on a single nanomechanical mode (the fundamental mode of one beam half with frequency $\Omega_\mathrm{m}$, sketched in green in Fig.~\ref{fig:Figure1}b) such that it responds as a digital logic gate. To this end, we seek a step-like nonlinear relation between mechanical output amplitude and applied input force, allowing low and high amplitudes on either side of the step to map naturally to binary states. Such nonlinearity is crucial to digital logic because it provides decision making (thresholding) as well as signal restoration (error correction, enabling cascading into larger computational networks~\cite{jin2024engineering}). 
Input information is encoded in the intensity of a `signal' laser, modulated at a frequency $\Omega_\mathrm{d}$ near $\Omega_\mathrm{m}$. Tuned to the nanocavity resonance, this laser exerts a radiation-pressure force $f(t)$ that drives the mechanical mode of interest. The induced response, the output signal $x(t)$, is recorded as a small modulation on a far-detuned `readout' laser. 
Crucially, we introduce a separate `function' laser that programs the resonator's nonlinear dynamical response. Switching this laser on or off yields dramatically different responses of the nanobeam to the input signal $f(t)$, as shown in the left panels of Fig.~\ref{fig:Figure1}c,d: With the function laser on, the mechanical motion responds strongly nonlinearly to the modulated signal laser force $f(t)$. Figure~\ref{fig:Figure1}c (left top) evidences a shark-fin-shaped, hysteretic frequency response versus modulation frequency $\Omega_\mathrm{d}$ --- the hallmark of Duffing nonlinearity~\cite{schmid2016fundamentals}. Without the function laser, the response is Lorentzian, i.e., fully linear (Fig.~\ref{fig:Figure1}d), emphasizing the optical origin and tunability of the nonlinearity.

This nonlinear control arises from a strong, \emph{nonlinear optical spring effect} induced by the function laser~\cite{Slim2026strong}. 
Because the cavity length depends on the nanobeam displacement $x$, the optical force $f_{\mathrm{opt}}$ experienced by a resonator depends on its displacement --- a mechanism known as cavity optomechanical backaction~\cite{aspelmeyer2014cavity}. The nanobeam equation of motion (EOM) thus reads
\begin{equation}
    \ddot{x}=-\Omega_\mathrm{m}^{2} x -\Gamma\dot{x}+ f_{\mathrm{opt}}(x),
\label{eq:Eq2}
\end{equation}
where $\Gamma$ is the mechanical damping rate and $m$ the resonator's effective mass, which we also use to normalize $f_\mathrm{opt}$. 
As detailed below, $f_{\mathrm{opt}}(x)$ is remarkably nonlinear in $x$; indeed, it follows a Lorentzian associated with the cavity response, as sketched in Fig.~\ref{fig:Figure1}b. Expanding this force in a Taylor series around the equilibrium position, the EOM can be rewritten as
\begin{equation}
\ddot{x} \approx - (\Omega_\mathrm{m}+\delta\Omega)^2x - \Gamma\dot{x} -\beta x^3 + f(t),
\label{eq:Eq1}
\end{equation}
where $\delta\Omega$ is the regular linear optical spring shift~\cite{aspelmeyer2014cavity}, and we have absorbed a static displacement shift into the definition of $x$ and neglected possible $x^2$ terms (see Methods). Equation~\ref{eq:Eq1} is the canonical equation of motion of a Duffing oscillator, with nonlinear strength parametrized by coefficient $\beta$, which is proportional to the third derivative $\partial^3 f_\mathrm{opt}/\partial x^3$. This induces a pronounced nonlinear response of $x$ to any additional force $f(t)$, such as that from the signal laser~\cite{Slim2026strong}.

For a specific modulation detuning $\delta\Omega_\mathbf{d} =\Omega_\mathrm{d}-(\Omega_\mathrm{m}+\delta\Omega)$, this strong Duffing nonlinearity produces a discontinuous jump in oscillation amplitude as the signal laser modulation strength is increased, as shown in Fig.~\ref{fig:Figure1}c. This provides precisely the step-like response required for decision making and level restoration in digital logic, with output states more sharply defined than inputs. The low and high mechanical amplitudes represent bits ``0'' and ``1'', respectively. We use these to implement an OR gate, illustrated in Fig.~\ref{fig:Figure1}c (right panel). The gate receives two inputs, each encoded by an independent signal laser: 
a logical ``1'' (``0'') corresponds to a high (low) modulation strength of the respective laser. When both inputs are ``0'', the resonator remains on the low-amplitude branch (output ``0''). If at least one input is ``1'', the oscillator switches to the high-amplitude branch (output ``1''), consistent with a digital OR gate.
Further discussion of the precise response function and logic requirements is provided in the Supplementary Information.

\subsection*{Optical reconfiguration of computing function}
%TC:ignore
\begin{figure*}[t]
    \centering
    \includegraphics[width=0.9\linewidth]{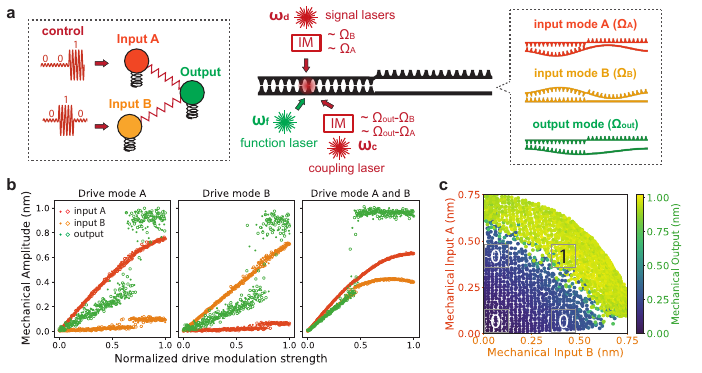}
    \caption{{\bf Cascadable logic gates using multiple mechanical eigenmodes.} {\bf a.} Schematic illustration of cascadable logic gates. Three mechanical eigenmodes of the nanobeam are employed, with two modes serving as input modes A and B and the third mode as the output mode. In addition to the signal lasers and the function laser used in the single-mode computation, an extra coupling laser is applied to establish connections between the input and output modes. The coupling laser is intensity-modulated near both $\Omega_\mathrm{out}-\Omega_\mathrm{A}$ and $\Omega_\mathrm{out}-\Omega_\mathrm{B}$. {\bf b.} The mechanical oscillation amplitudes of the three eigenmodes when driving input mode A only, input mode B only, and both input modes A and B using an intensity modulated signal laser. The drive modulation strength is normalized, for driving mode A (B), the maximum modulation amplitude corresponds to 0.42~V (0.24~V). The saturation in the right panel arises from the laser intensity modulation becoming more nonlinear at higher modulation depths. {\bf c.} Experimental demonstration of an AND gate, where both input and output states are represented by the mechanical oscillation amplitude.}
    \label{fig:Figure3}
\end{figure*}
%TC:endignore

Both the strength and character of the nonlinearity are controlled through the function laser. 
The optical force is
\begin{align}
    f_\mathrm{opt}= \frac{\hbar Gn_\mathrm{max}(t)}{m}\frac{1}{1+4(\omega-\omega_\mathrm{cav}+Gx)^2/\kappa^2},
\label{eq:Eq3}
\end{align}
with $\omega$ the laser frequency,  $\omega_\mathrm{cav}$ the cavity resonance frequency, $\hbar$ the reduced Planck constant, and $n_\mathrm{max}$ the maximum intracavity photon number for a resonant laser (see Supplementary Information). 
The force follows a Lorentzian dependence on the detuning $\omega - \omega_\mathrm{cav} + Gx$, yielding 
the effective Duffing nonlinearity
\begin{equation}
     \beta = \frac{\hbar G n_\text{max}}{m}\frac{G^3}{\kappa^3}\frac{32u_0(u_0^2-1)}{(1+u_0^2)^4},
\end{equation}
where $u_0=2(\omega-\omega_\mathrm{cav})/\kappa$ is the normalized optical detuning.
In the experiments above, we tuned the function laser frequency $\omega_\mathrm{f}$ to the red-detuned maximum of this nonlinearity, realizing a strong softening Duffing nonlinearity; hardening can be achieved at other frequencies (see  \EDF~\ref{fig:Ext2}).

Beyond the optical frequency, the nonlinear response is also controlled by the modulation frequency detuning from mechanical resonance $\delta\Omega_\mathrm{d}=\Omega_\mathrm{d}-(\Omega_\mathrm{m}+\delta\Omega)$. 
To satisfy the cascading constraint discussed above, we choose a modulation detuning where hysteresis between forward and backward sweeps of modulation strength is small (Fig.~\ref{fig:Figure1}c and \EDF~\ref{fig:Ext2}c). This suppresses memory effects, ensuring that outputs depend only on current inputs rather than previous states. The logic gates can therefore operate under continuous signals without resetting the system to the ``00'' state before each input sequence. Nevertheless, Fig.~\ref{fig:Figure1}c shows a pronounced response step, even larger than expected for an ideal Duffing nonlinearity. This strong level separation benefits computing and indicates contributions beyond the idealized Duffing model (see Supplementary Information).

Laser control allows the logic gate function to be reconfigured in real time by adjusting optical power and wavelength. This programmability avoids the need to adjust device design and redefine digital thresholds for inputs and/or outputs to change logic function~\cite{song2019additively, raney2016stable, byun2024integrated, el2021digital, mei2021mechanical,guerra2010noise, fredkin2014, mahboob2011interconnect}. 
We realize all six fundamental logic gates in the same device, while keeping the digital thresholds for both inputs (laser intensity modulation strengths) and output (mechanical oscillation amplitude) unchanged, as shown in Fig.~\ref{fig:Figure2}.  
We switch between AND and OR operations by adjusting the signal laser wavelength, which primarily modifies the driving force on the nanobeam while also affecting the Duffing nonlinearity. 
To realize NOR and NAND operations, the nanobeam must reach the bit ``1'' output state even for inputs ``00''. We therefore introduce an oscillatory bias modulation on the intensity modulator, applied out of phase with the original modulation. For inputs ``11'', this bias cancels the input modulation, driving the nanobeam to the bit ``0'' state (low-amplitude branch). In this way, the AND (OR) gate can be converted into NOR (NAND) operation. With a similar approach at half the bias strength, an OR gate can be reconfigured into an XNOR gate. Finally, by adjusting the relative phase between the two input signals to be out of phase, we realize the XOR gate.

%TC:ignore
\begin{figure*}[]
    \centering
    \includegraphics[width=0.75\linewidth]{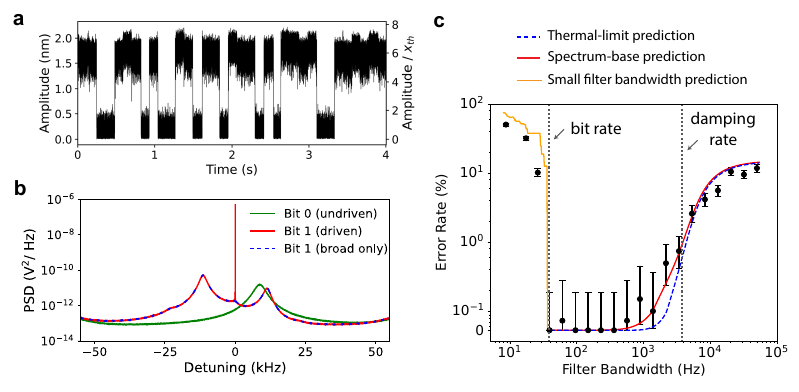}
    \caption{{\bf Computation fidelity near the thermal limit.} {\bf a.} Time trace of an OR gate output under a continuous input sequence. Since inputs 01 (10) and 11 both produce output bit 1, the output exhibits two mean levels. Data are acquired with the lock-in amplifier filter bandwidth set to 10 kHz. The mechanical oscillation amplitude is presented both in absolute units and normalized by the thermal displacement $x_\mathbf{th}$, providing a dimensionless measure relative to thermal motion. {\bf b.} Power spectral density of the output signal in the undriven state (bit 0) and the driven state (bit 1), where the driven case corresponds to input 11. For the driven case, the spectrum exhibits two prominent Lorentzian peaks. The left peak corresponds to the shifted effective eigenfrequency resulting from the softening nonlinearity, while the right peak arises from nonlinear mixing between this Lorentzian response and the $\delta$ function peak. {\bf c.} Predicted error rate of the OR gate. Black scatter points and error bars show the experimental data and its confidence interval. The green curve is the prediction based on the thermal-noise limit inferred from the undriven data, while the red curve is based on the power spectral density of the driven data. When the filter bandwidth is smaller than the operation bit rate, the readout effectively averages over the last few bits, and the corresponding prediction is shown in yellow. The bit rate here is approximately 40 Hz.}
    \label{fig:Figure4}
\end{figure*}
%TC:endignore

\subsection*{Cascadable logic}
For both analog and digital computing platforms, complex functionality requires interconnecting multiple elements. 
Here, we show how to program connections between nonlinear gates and other mechanical resonators.
We use multiple degrees of freedom within a single nanobeam, exploiting its mode spectrum as a synthetic dimension (Fig.~\ref{fig:Figure3}a)~\cite{mathew2020synthetic,delpino2022}. 
As a demonstration, we implement an AND gate using three mechanical eigenmodes: two serving as inputs and one as output (Fig.~\ref{fig:Figure3}a). 
To enable information flow among the three mechanical eigenmodes, we apply an additional detuned `coupling' laser. By modulating its intensity, and thus the spring effect it induces, near the frequency difference between two mechanical modes, this laser can couple any pair of eigenmodes~\cite{mathew2020synthetic}. The coupling strength is set by the wavelength, power, and modulation depth of the laser, enabling real-time optical control of network connectivity.

We choose as output a mode showing strong Duffing nonlinearity, that is coupled through two simultaneous coupling modulations to two input mechanical modes A and B with smaller nonlinearity (due to higher effective mass, Fig.~\ref{fig:Figure3}b). 
We use this configuration to implement a fully in-mechanics AND logic gate, as evidenced in Fig.~\ref{fig:Figure3}c, where this time both inputs are defined as the mechanical oscillation amplitudes of modes A and B. 

Interestingly, these experiments reveal a weak backaction effect associated with the Duffing-based resonator logic. As the left two panels of Fig.~\ref{fig:Figure3}b show, the undriven input modes are also weakly excited as the output switches to the ``1'' state. 
This arises from backaction through coupling between the output mode and the undriven input mode. Owing to nonreciprocity induced by the strong Duffing nonlinearity, this backaction remains weak: information transfer from the input modes to the output is much stronger than transfer from the output mode back to the input modes, enabling predominantly unidirectional information flow.

\subsection*{Thermally limited computation fidelity}
Importantly, we recognize in Fig.~\ref{fig:Figure3}c that the 3-mode AND gate operates at mechanical input amplitudes of only 0.4~nm, reflecting the exceptionally strong nonlinearity and efficient couplings enabled by optomechanical interactions.
In fact, these logic gates operate close to the intrinsic thermal fluctuation amplitude $x_\mathrm{th}=\sqrt{k_BT/(m_\mathrm{eff}\Omega^2)}$ of the resonators, enabling exploration of computation in the regime where its performance is governed by fundamental thermodynamic noise. 
Figure~\ref{fig:Figure4}a shows the output (demodulated oscillation amplitude) time trace of an OR gate under a continuous input sequence. Two distinct bit-``1'' levels appear at $\sim6$ times $x_\mathrm{th}$, corresponding to the input states 01(10) and 11. The substantial noise on these levels and on the bit-``0'' traces is of thermal origin, as supported by their power spectral densities (PSD) 
shown in Fig.~\ref{fig:Figure4}b. For state ``0'', the PSD shows the Lorentzian spectrum of thermomechanical fluctuations, slightly detuned from the drive. For state ``1'', the main fluctuation peak switches to opposite detuning and additional thermal noise harmonics appear because of nonlinear transduction of the driven motion~\cite{leijssen2017nonlinear}. 

As a result, thermal fluctuations also impact gate fidelity. Figure~\ref{fig:Figure4}c shows the measured error rate of an OR gate, obtained from thousands of bit operations, as a function of the filter bandwidth applied to the demodulated output. This bandwidth must generally exceed the bit rate; otherwise, the readout averages over several preceding bits, resulting in high error rate (confirmed by a numerical model, yellow curve). 
For higher bandwidths, 
faster output state evaluation increases the computation error rate. This becomes particularly pronounced when the filter bandwidth approaches the modulation-detuning frequency seen in Fig.~\ref{fig:Figure4}b. 
Beyond that bandwidth, most thermal noise is included in the measurement, dominating the observed error rate. 

To identify the noise sources, we independently predict error rates from the measured PSD in Fig.~\ref{fig:Figure4}b. The red solid line in Fig.~\ref{fig:Figure4}c includes all fluctuations in the PSD, assuming they are Gaussian distributed (see Supplementary Information). 
Besides thermal noise, these may include $1/f$ technical noise; i.e., low-frequency phase noise near the sharp peak at zero detuning~\cite{bachtold2022mesoscopic}. For bandwidths above a few kHz, however, the measured error rate follows 
the model based on only the thermal noise observed in Fig.~\ref{fig:Figure4}b, shown as blue dashed line in Fig.~\ref{fig:Figure4}c. The latter is estimated by excluding noise near the driving peak (see Methods and Supplementary Information). 
The correspondence holds specifically 
for bandwidths on the order of the mechanical damping rate, which sets the intrinsic dynamical response timescale of the mechanical system; If the bit rate would exceed this rate, the nanobeam would not have sufficient time to reach the desired state before the next input. We therefore conclude that when the logic gate operates at its natural fastest speed, its error rate is dominated by noise of intrinsic thermal origin.

\subsection*{Outlook}

In conclusion, we demonstrated physical computation in a nano-optomechanical system, where cavity radiation-pressure backaction induces a strong nonlinear response and laser wavelength and power program computational function in real time. 
This approach is not only flexible but also yields remarkably strong nonlinearities, enabling nanomechanical logic operations close to intrinsic thermal amplitudes. With realistic device-parameter improvements~\cite{leijssen2017nonlinear}, this amplitude could be reduced far below $x_\mathrm{th}$. Although in such a regime our proof-of-concept demonstration of digital logic would cease to be relevant, it could enable the use of noise as a computational resource, e.g. for probabilistic computing~\cite{Camsari2019p-bits}. More generally, we emphasize that the methods presented here are by no means restricted to digital logic. Instead, we envision that deep control over nonlinearities and network connectivity will benefit analog computing paradigms including physical neural networks, where reconfiguration is essential for in-material learning. Interestingly, optomechanical coupling can control phenomena such as nonreciprocity and non-Hermiticity in resonator networks~\cite{delpino2022}, offering prospects to uncover fundamental relations between computational performance and network symmetries and topologies. Because the demonstrated mechanisms apply to any cavity optomechanical platform, they open the way to optomechanically programmed computing across scales and integration platforms. Combined with the highly sensitive, all-site readout and actuation capabilities of optomechanical systems, this establishes cavity optomechanics as a powerful platform for physical computing.

%TC:ignore

\clearpage
\setcounter{section}{0}
\setcounter{equation}{0}
\renewcommand{\theequation}{A\arabic{equation}}%
\setcounter{figure}{0}
\renewcommand{\thefigure}{A\arabic{figure}}%
\renewcommand{\figurename}{\EDF}%
\renewcommand\appendixname{}
\subsection*{METHODS}

\section{Sample and Experimental Setup}
\subsection{Sample Design and Fabrication}
The suspended silicon nanobeam used in this study is fabricated on a silicon-on-insulator chip in a 220~nm thick silicon layer patterned with a sliced photonic crystal nanocavity through electron beam lithography and reactive ion etching. For details on design and nanofabrication, see Ref.~\cite{delpino2022}. The cavity hosts an optical mode with a resonance frequency of $\omega _0/ (2\pi) =195.7$ THz and a linewidth of $\kappa / (2\pi) = 320$ GHz. This optical mode is coupled to several flexural mechanical modes of both halves of the nanobeam. We specifically use three mechanical modes, i.e. those with the second to the fourth lowest frequencies, with resonance frequencies $\Omega_{\mathrm{m},i} / 2\pi = {5.09, 12.51, 18.03}$~MHz, as shown in \EDF~\ref{fig:Ext1}a. The linewidths of these modes are approximately $\Gamma_i / 2\pi \approx 2-5~$kHz. 
\subsection{Experimental Layout}
\label{sec:setup}
The optical setup layout is schematically depicted in \EDF~\ref{fig:Ext1}b. Four lasers are employed in the experiments. Modulated Laser A and Modulated Laser B are modulated by intensity modulators (IM, CovegaMach-10 056 and LN81-10, respectively). The modulator for Modulated Laser 1 is followed by a polarization controller (PC) since this modulator was not polarization maintaining. Each modulator has two inputs: the alternating current (AC) component is provided by the lock-in amplifier (LIA, Zurich Instruments UHFLI), while the direct current (DC) component is supplied by a function generator (FG). 
In the single mechanical mode experiments (Fig.~\ref{fig:Figure1} and Fig.~\ref{fig:Figure2} in the Main Text), two Modulated Lasers are used as input signal lasers. In the multiple-mechanical modes experiments (Fig.~\ref{fig:Figure3} in the Main Text), one Modulated Laser is used as the input signal, modulated by two mechanical frequencies provided by the lock-in amplifier, while the other Modulated Laser serves as the coupling laser. 
The outputs of the four lasers are combined using three beam splitters (BS). The power-splitting ratios of the beam splitters are selected to balance the optical powers of the four lasers before coupling to free space through a collimator (COL).
The laser beam is focused at normal incidence onto the sample surface. The sample is placed in a vacuum chamber and aligned with the laser focus using a piezo-actuated precision stage. The nanobeams are aligned at an angle of 45 degrees to the incident polarization. A small fraction (of the order of 1\%) of the incident light couples to the nanocavity. Since the reradiated light exiting the nanobeam and collected by the same lens has a different polarization than the incident beam, a portion of this light is transmitted through the polarizing beam splitter (PBS). After fiber coupling, a bandpass filter (BPF) selectively filters the frequency of the detection laser which is then detected on a low-noise photodiode (PD) and analyzed using the LIA.

\subsection{Laser Sources and Power}
All lasers used in this paper are Toptica CTL models, operated at their maximum power of 25mW-30mW at the source. After passing through the fiber beam splitters and emerging into free space from the launch collimator, the function and detection lasers have $\sim$1-2 mW power each, and the modulated lasers have $\sim$0.2 mW power per laser beam. Approximately 70\% of the power is preserved just before the polarizing beam splitter located ahead of the vacuum chamber. At the collection collimator in front of the photodiode, $\sim$20\% of the original power remains in free space, and only $\sim$3-5\% is coupled back into the fiber. The filtered detection laser power at the photodiode is $\sim$20-30 $\mu$W.

\subsection{Laser Intensity Modulation}
\label{sec:intensity modulators}
We encode the input information by modulating the laser intensity using electro-optic fiber modulators. The modulation strength of each modulator is controlled by a combination of an AC signal and a DC bias. 
We calibrate both types of modulators to specific DC bias values within a suitable AC range to operate in their linear range and suppress the second and third order harmonics. This has little impact on single-mode experiments, but in multimode experiments, where one laser is driven by a single modulator at two modulation frequencies, higher order harmonics and intermodulation products can unintentionally couple to other modes if they are insufficiently suppressed. 
By scanning the full AC-DC parameter space, we identified the optimal operating points: Modulator A at a DC bias of 2.0 V, and Modulator B at a DC bias of 3.3 V. As shown in Fig.~\fupdate{S12} in Supplementary Information, the higher-order harmonics are negligible. In most experiments, we use an AC range of 0-0.1 V, over which the modulators are approximately linear.

\section{Optomechanical theory}
\subsection{Nonlinear optical spring effect and driving}
\label{sec:theory_singlemode}
We briefly introduced the equation of motion (EOM) of the mechanical oscillator in Eq.~\ref{eq:Eq2} and Eq.~\ref{eq:Eq3} of the Main Text. In this section, we provide a more detailed theory, in line with~\cite{Slim2026strong}. Substituting Eq.~\ref{eq:Eq3} into Eq.~\ref{eq:Eq2} gives
\begin{equation}
    \ddot{x}=-\Omega_\mathrm{m}^{2} x -\Gamma\dot{x}+ \frac{\hbar G n_\mathrm{max}}{m}\frac{1}{1+4(\omega-\omega_\mathrm{cav}+Gx)^2/\kappa^2}.
\label{eq:EOM_single}
\end{equation}
Expanding the last term in a Taylor series up to third order in $x$, we obtain
\begin{align}
    &\ddot{x} = -\alpha x - \mu x^2 - \beta x^3 -\Gamma\dot{x} + f,\nonumber\\
    &\alpha = \Omega_\text{m}^2+\frac{\hbar G n_\text{max}}{m}\frac{G}{\kappa}\frac{4u_0}{(1+u_0^2)^2},\nonumber\\
    & \mu = \frac{\hbar G n_\text{max}}{m}\frac{G^2}{\kappa^2}\frac{4(1-3u_0^2)}{(1+u_0^2)^3},\nonumber\\
    & \beta = \frac{\hbar G n_\text{max}}{m}\frac{G^3}{\kappa^3}\frac{32u_0(u_0^2-1)}{(1+u_0^2)^4}, \nonumber\\
    & f = \frac{\hbar G n_\text{max}}{m}\frac{1}{1+u_0^2},
\label{eq:coefficients}
\end{align}
where $u_0= 2(\omega-\omega_\text{cav})/\kappa$ is the normalized optical detuning. To illustrate more intuitively how the optical detuning affects these coefficients, we plot the normalized coefficients in \EDF~\ref{fig:Ext1}c. The optically induced linear term proportional to $\alpha$ results in a shift of the effective mechanical eigenfrequency $\delta\Omega$, commonly referred to as the optical spring effect~\cite{aspelmeyer2014cavity},
\begin{equation}
\delta\Omega\approx\frac{g_0^2n_\mathrm{max}}{\kappa}\frac{4u_0}{(1+u_0^2)^2}.
\label{eq:springshift}
\end{equation}
The result of the optical spring effect on the thermomechanical fluctuation spectrum of one of the mechanical modes is illustrated in \EDF~\ref{fig:Ext1}d as a function of the function laser frequency while reading out the displacement fluctuations with the readout laser kept at fixed detuning.

The function laser introduces various nonlinearities, including specifically the Duffing nonlinearity with strength proportional to $\beta$. By modulating the signal laser intensity, we modulate $n_\mathrm{max}$ with a sinusoidal function, thereby generating a periodic driving force through the term $f$. The choice of laser wavelength is described in Sec.~\ref{sec:calibration}. Further details of the theory are provided in the Supplementary Information.

\subsection{Optically coupled mechanical eigenmodes}
In the previous section, we considered only a single eigenmode of the nanobeam. In reality, however, the nanobeam exhibits multiple higher-order eigenmodes, so the full equation of motion (EOM) is:
\begin{align}
    \ddot{x}_i=& -\Omega_{i}^{2} x_i -\Gamma_i\dot{x}_i\nonumber\\
    &+\frac{\hbar G_i n_\mathrm{max}(t)}{m}\frac{1}{1+4(\omega-\omega_\mathrm{cav}+\sum_{j}G_jx_j)^2/\kappa^2}.
\end{align}
For terms with $j\neq i$, the motion occurs at different eigenfrequencies. After transforming to the rotating frame, these contributions oscillate rapidly. They can therefore be neglected under the rotating-wave approximation (RWA). This justifies the single-mode treatment of the nanobeam and allows the EOM to be reduced to Eq.~\ref{eq:EOM_single}.

If, however, an additional laser is introduced whose intensity is modulated at the frequency difference between two eigenmodes, the two modes can be coherently coupled through this driving field~\cite{mathew2020synthetic}, as we describe in the following. For simplicity, we consider here the coupling term to be linear, by shifting the displacement to the new equilibrium state and taking the Taylor expansion to the first order, to obtain:
\begin{align}
    \ddot{x}_i=& -\Omega_{i}^{2} x_i -\Gamma_i\dot{x}_i\nonumber\\
    &-\frac{\hbar G_i n_\mathrm{max}(t)}{m\kappa}\frac{4u_0}{(1+u_0^2)^2}\sum_jG_jx_j.
\end{align}

Modes $i$ and $j$ can be coupled using a coupling laser whose intensity is modulated periodically at the frequency $\Omega_j - \Omega_i + \delta\Omega_{ij}$.
In this case, $n_\mathrm{max}(t)$ becomes time dependent:
\begin{equation}
    n_\mathrm{max}(t) = \bar{n}_\mathrm{max} f_0 e^{i(\Omega_j-\Omega_i+\delta\Omega_{ij})t},
\end{equation}
where $\bar{n}_\mathrm{max}$ denotes the maximum intracavity photon number, and $f_0$ characterizes the modulation strength. We move to the rotating frame by defining
\begin{align}
    \tilde{x}_i = x_i e^{-i\Omega_i t}.
\end{align}
Applying the rotating-wave approximation (RWA) and the slowly varying envelope approximation (SVEA), we obtain in the frequency domain
\begin{align}
    &\tilde{x}_i(\omega) = \chi_i(\omega)\,G_{ij}\,\tilde{x}_j(\omega-\delta\Omega_{ij}),\\
    &\chi_i(\omega) = \frac{1}{\omega +i \Gamma_i/2},\\
    & G_{ij} = \frac{\hbar G_i G_j \bar{n}_\mathrm{max} f_0}{2\Omega_i m\kappa}
    \frac{4u_0}{(1+u_0^2)^2}, \label{eq:coupling strength}
\end{align}
where $\chi_i(\omega)$ is the mechanical susceptibility, and $G_{ij}$ is the optomechanical coupling coefficient, which depends on both the optical detuning $u_0$ and the modulation strength $f_0$.

\section{Calibration of Experimental Control Parameters}
\label{sec:calibration}
To account for small drifts in alignment that lead to variations in the optical coupling efficiency, the frequent calibration procedures are performed to determine the appropriate selection of control parameters during each experiment. 
This includes a modulation calibration including mechanical detuning and phase for the signal and coupling lasers whose intensities are modulated, and an optical calibration to set laser wavelength and power. 
The calibrations are carried out according to the procedures described below.

\subsection{Mechanical modulation detuning $\delta\Omega_{\text{d}}$}
We define the mechanical drive detuning as $\delta\Omega_{\text{d}}=\Omega_{\text{d}}-(\Omega_{\text{m}}+\delta\Omega)$, where $\Omega_{\text{m}}$ is the natural mechanical frequency and $\delta\Omega$ is the optomechanical spring shift (Eq.~\ref{eq:springshift}).
For calibration, we enable all lasers that will be used for the experiments without modulation and measure the thermal power spectrum density (PSD). We fit the PSD to obtain the effective mechanical resonances $\Omega_{m,k}+\delta\Omega_{k}$ for every mode $k$, which we then use to compute $\delta\Omega_{\text{d,k}}$.

To achieve the logic gate, we need to select a specific mechanical detuning $\delta\Omega_{\text{d}}$. To this end, we sweep the modulation detuning, and for each detuning value, we vary the modulation strength while recording the signal from the photodetector. When the absolute value of the detuning is large, the system exhibits strong hysteresis (\EDF~\ref{fig:Ext2}c left). Conversely, if the detuning is too small, no multistability region appears and thus no “jump” is observed. Neither of the above two cases satisfies the cascading constraint required for digital computation (see Supplementary for details). Therefore, we select the modulation detuning at the boundary where the system transitions from no hysteresis to the onset of hysteresis (\EDF~\ref{fig:Ext2}c right).

\subsection{Modulation Phase}
For implementing logic gates, we must choose the appropriate modulation phase for the two independent signal(drive) lasers. As shown in Fig.~\ref{fig:Figure2} in the Main Text, the two drives are selected to be either in phase or out of phase, depending on the specific logic gate type. Because the two drives use different cables and modulators, the simplest and most reliable procedure is to measure the mechanical response while sweeping the relative phase. When the mechanical amplitude is maximized (minimized), the two drives are in phase (out of phase), respectively.

\subsection{Optical detuning $\Delta_0$}
\label{sec:cali_optical detuning}

The detection laser wavelength is kept fixed at the far-detuned wavelength of 1540~nm. For other lasers, we define, for each laser $j$, a frequency $\omega_j$ with the corresponding optical detuning $\Delta_{0,j}=\omega_j-\omega_\text{cav}$. The wavelengths of these lasers are calibrated before each experimental run to ensure accuracy.
The signal laser frequency $\omega_\mathrm{s}$ is set to cavity resonance $\omega_\mathrm{cav}$ to maximize the driving force while avoiding additional Duffing nonlinearity (Fig.~\ref{fig:Figure1}b).

The function laser is used to induce optical nonlinearity. To maximize the nonlinear coefficient $\beta$ in Eq.~\ref{eq:coefficients}, we choose the normalized optical detuning to be $u_\mathrm{f}=\pm\sqrt{1-\frac{2}{5}\sqrt{5}}$. We operate in the softening regime, corresponding to $u_\mathrm{f}=\sqrt{1-\frac{2}{5}\sqrt{5}}$. The signal laser is set at zero detuning, $u_\mathrm{s}=0$, to avoid interference with the function laser and to maximize the driving strength. The coupling laser is set at $u_\mathrm{c} = 1/\sqrt{3}$ to maximize the coupling strength (Eq.~\ref{eq:coupling strength}).

\begin{figure*}
    \centering
    \includegraphics[width=\linewidth]{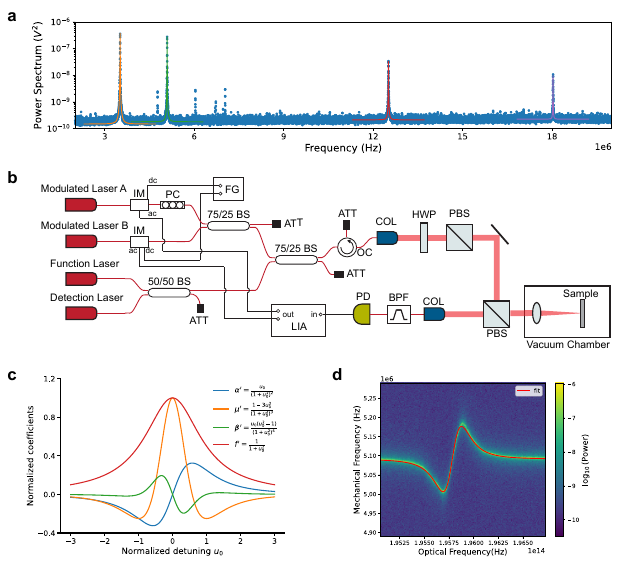}
    \caption{ {\bf Nanobeam optomechanical system.}  {\bf a.} Nanobeam power spectrum measured with a far-detuned detection laser. The first four mechanical eigenmodes are observed at resonance frequencies $\Omega_{\mathrm{m},i}/2\pi=\{3.51,\,5.09,\,12.51,\,18.03\}$~MHz, with corresponding linewidths $\Gamma_i/2\pi=\{2.93,\,2.14,\,5.13,\,4.95\}$~kHz. {\bf b.} Schematic of the experimental setup. IM, intensity modulator; PC, polarization controller; FG, function generator; BS, beam splitter; ATT, attenuator; OC, optical circulator; COL, collimator; HWP, half waveplate; PBS, polarized beam splitter; BPF, bandpass filter; PD, photodiodes; LIA: lock-in amplifier. {\bf c.} Normalized optomechanical coefficients based on Eq.~\ref{eq:coefficients}, where $\alpha'$, $\mu'$, $\beta'$ and $f'$ correspond to the coefficients of the linear, quadratic, cubic, and constant terms in $x$ in the EOM, respectively. {\bf d.} Optomechanical linear spring effect, as observed on the thermomechanical fluctuation spectra while the function laser frequency is swept. The red curve shows a fit of Eq.~\ref{eq:springshift} to the Lorentzian peak frequencies.}
    \label{fig:Ext1}
\end{figure*}

To determine the laser wavelengths corresponding to the desired normalized optical detuning $u_0$, we sweep the laser wavelength and measure the PSD, fitting Lorentzians to the spectra for each optical wavelength $\omega_l$ and extracting the effective mechanical resonance frequency. We use Eq.~\ref{eq:springshift} to fit the effective mechanical frequency curve (solid red line) shown in \EDF~\ref{fig:Ext1}d), from which we determine the cavity resonance frequency $\omega_\mathrm{cav}$, corresponding to $u_0 = 0$. Based on this fit, we can also determine the wavelengths of the other lasers according to their defined $u_0$. 
Static shifts of the cavity resonance frequency $\omega_{\text{cav}}$ due to the intracavity intensity can be neglected, as evidenced by the symmetric optical spring response shown in \EDF~\ref{fig:Ext1}d. We therefore perform a single calibration and keep $\omega_{\text{cav}}$ fixed throughout.

\subsection{Optical laser power}
All lasers are operated at or near their maximum output by setting the laser current to 300 mA. The only exception are the two signal lasers used to generate the two inputs of the logic gate. Because the two modulators have different characteristics, the powers of these two lasers were calibrated to ensure an equal contribution to the gate operation. The calibration was performed by maintaining the laser powers as high as possible while adjusting them until the measured outputs were equal.

\section{Data Acquisition and Processing}
\paragraph{\textbf{Lock-in amplifier.}}
All data in this paper are acquired with the lock-in amplifier. The signal from the photodiode is fed into a lock-in amplifier which demodulates the signal. The demodulation creates a complex two quadratures, X (in-phase) and Y (out-of-phase), which are passed through a low-pass filter with a configurable bandwidth. The output is a complex signal $V = X + iY$ of which we record timetraces. We can transform a timetrace of the output $V(t)$ into a power spectral density by calculating the absolute value squared of its Fourier transform. Because the spectral shape of the low-pass filter is known, we can negate its effect on the spectrum. 
Unless noted otherwise, we record time traces of the lock-in output after setting a 1~kHz demodulation bandwidth. This yields practical acquisition times while avoiding excessive thermal noise.
\paragraph{\textbf{Single-mode duffing nonlinearity.}}
For the data in Fig.~\ref{fig:Figure1}c and d, \EDF~\ref{fig:Ext2} and \fupdate{Fig.~S15} in Supplementary Information, the measurements are performed by continuously sweeping either the modulation detuning or the modulation strength in both forward and backward directions, without turning off the lasers between sweeps. This protocol reveals the hysteresis. For each detuning-strength pair, we record a single data point, represented as a complex number that contains both amplitude and phase.

For the data in \EDF~\ref{fig:Ext2} and Fig.~\fupdate{S15} in Supplementary Information, each set of control parameters is measured 10 times, and the results are overlaid in the plots. For the data shown in Fig.~\ref{fig:Figure1} in the Main Text, each measurement is repeated 50 times. For clarity, only one representative data set is displayed in each plot. The complete set of 50 repetitions is shown in Fig.~\fupdate{S16} in Supplementary Information. 

\paragraph{\textbf{Logic gates.}} 
As shown in Fig.~\ref{fig:Figure1}, Fig.~\ref{fig:Figure2} and Fig.~\ref{fig:Figure3} of the Main Text, we sweep the modulation strengths of two signal lasers. To demonstrate continuous, reset-free operation of the logic gate, the lasers remain on throughout each measurement set. The sequence of drive combinations is randomly permuted using Python’s "shuffle" function. Each two-dimensional sweep in Fig.~\ref{fig:Figure1} is repeated 10 times, with an independently shuffled sequence for each repeat. In Fig.~\ref{fig:Figure1} in the Main Text we display one representative sweep. The mean amplitude and phase from ten measurements, along with their standard deviations, are presented in \fupdate{Fig.~S13 and Fig.~S14} in the Supplementary Information.

\paragraph{\textbf{Mechanical oscillation amplitude calibration.}}

As shown in Fig.~\ref{fig:Figure3}c,d and Fig.~\ref{fig:Figure4}a, the mechanical oscillation amplitude is presented in absolute units and normalized to the thermal displacement $x_{\mathrm{th}}$. In the following, we detail the procedure used to convert the output RF signal (in volts) into mechanical oscillation amplitude (in nm), as well as the determination of $x_{\mathrm{th}}$.

First, the thermal displacement used in Fig.~\ref{fig:Figure4}a is defined at $T = 300\,\mathrm{K}$ as
\begin{equation}
    x_{\mathrm{th}} = \sqrt{\frac{k_B T}{m_{\mathrm{eff}} \Omega^2}},
    \label{eq:x_th}
\end{equation}
where $k_B$ is the Boltzmann constant, $\Omega$ is the effective mechanical eigenfrequency including the optical spring shift $\delta \Omega$, i.e., $\Omega = \Omega_m + \delta \Omega$, and $m_{\mathrm{eff}}$ is the effective mass at the location of the optical cavity. The effective mass $m_{\mathrm{eff}}$ is calculated from COMSOL eigenmode simulations according to
\begin{equation}
    m_{\mathrm{eff}} = \frac{\int \rho(\mathbf{r}) |\mathbf{u}(\mathbf{r})|^2 dV} {|\mathbf{u}(\mathbf{r}_0)|^2
    },
\end{equation}
where $\rho(\mathbf{r})$ is the mass density, $\mathbf{u}(\mathbf{r})$ is the displacement field, and $\mathbf{r}_0$ denotes the position of the center of the optical cavity.

The mechanical oscillation amplitude $x(t)$ is proportional to the output signal $V(t)$ from the lock-in amplifier,
\begin{equation}
    x(t) = \frac{1}{aG} V(t),
    \label{eq:VtoX}
\end{equation}
where $G$ is the optomechanical coupling coefficient and $a$ accounts for additional conversion factors in the detection chain. To calibrate the factor $aG$, we use the integrated thermal noise spectrum:
\begin{align}
    \int_{0}^{+\infty} S_{vv}(f)\, df 
    &= \frac{a^2 G^2}{2\pi} \int_{-\infty}^{+\infty} S_{xx}(\omega)\, d\omega \nonumber\\
    &= a^2 G^2 \langle x^2 \rangle = \frac{a^2 G^2 k_B T_{\mathrm{eff}}}{m_{\mathrm{eff}} \Omega^2},
    \label{eq:Svv}
\end{align}
where $S_{vv}(f)$ is single-sided power spectral density, and  $T_{\mathrm{eff}}$ is the effective temperature arising from optomechanical backaction. It is given by
\begin{equation}
    T_{\mathrm{eff}} = \frac{\Gamma}{\Gamma_{\mathrm{tot}}} T,
    \label{eq:T_eff}
\end{equation}
where $\Gamma_{\mathrm{tot}}$ is the total mechanical damping measured with all lasers on, and $\Gamma$ is the intrinsic mechanical damping obtained from measurements with only a far-detuned detection laser.

Substituting Eqs.~\ref{eq:x_th}, \ref{eq:Svv}, \ref{eq:T_eff} into Eq.~\ref{eq:VtoX}, we obtain the calibrated displacement
\begin{equation}
    x(t) = \frac{V(t)}{ \sqrt{\frac{\Gamma_{\mathrm{tot}}}{\Gamma} \int_{0}^{+\infty} S_{vv}(f)\, df}}x_\mathrm{th}.
\end{equation}

\textit{\textbf{Data and code availability}}\\
All codes and data supporting this study are available in the public repository https://doi.org/10.5281/zenodo.20327075.

\textit{\textbf{Acknowledgments}}\\
We thank Daniel Koletzki for technical assistance. This work is part of the research programme of the Netherlands Organisation for Scientific Research. It is funded by the European Union, supported by ERC Grants 759644 (TOPP) and 101088055 (Q-MEME). Views and opinions expressed are however those of the authors only and do not necessarily reflect those of the European Union or the European Research Council Executive Agency. Neither the European Union nor the granting authority can be held responsible for them. The work was made possible in part by Quantum Delta NL, with funding from the National Growth Fund.

% \clearpage

\begin{figure*}[h!]
    \centering
    \includegraphics[width=\linewidth]{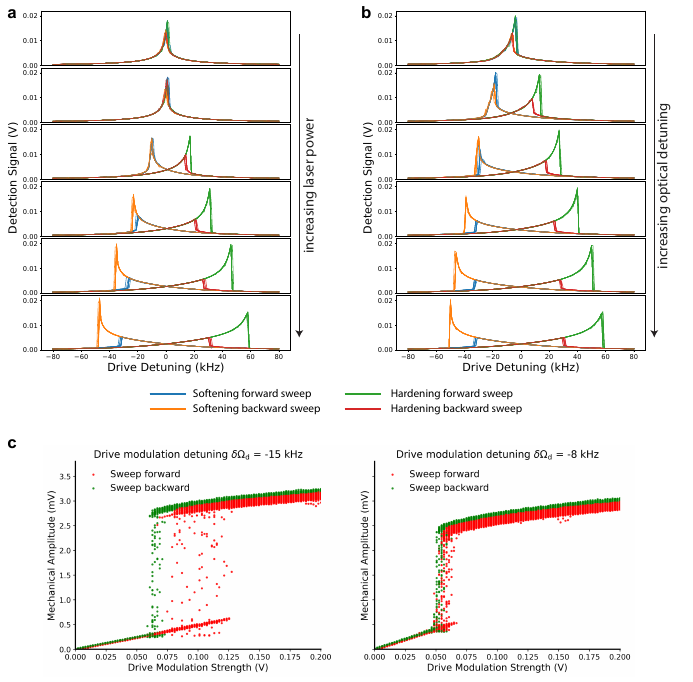}
    \caption{{\bf Programmable optomechanical nonlinearity.} {\bf a.} Tuning the nonlinearity by varying the function laser power. Rows from top to bottom correspond to increasing drive current. The function laser wavelength is set to maximize the Duffing coefficient. The function laser current from top to bottom are respectively set as $\{0.4, 60, 120, 180, 240, 300\}$ mA. {\bf b.} Tuning the nonlinearity by varying the function laser wavelength. Rows from top to bottom correspond to increasing absolute optical detuning $|u_0|$. The function laser current is set to 300 mA. For the softening Duffing, the function laser wavelength is tuned from top to bottom as $\{1532.39, 1532.31, 1532.23, 1532.14, 1532.06, 1531.98\}$ nm. For the hardening Duffing, the corresponding wavelengths are $\{1532.39, 1532.47, 1532.56, 1532.64, 1532.72, 1532.81\}$ nm. {\bf c.} Tuning the nonlinear response by varying the modulation frequency. Hysteresis appears when the drive modulation detuning $|\delta\Omega_\text{d}|$ is large.}
    \label{fig:Ext2}
\end{figure*}

\clearpage 
\onecolumngrid 
\setcounter{section}{0}
\setcounter{equation}{0}
\setcounter{figure}{0}

\renewcommand\appendixname{}

\newcommand{\mt}{of the Main Text}

\definecolor{darkgreen}{RGB}{105,150,150}
%Figure SX
\renewcommand{\figurename}{{\bf Figure }}
\renewcommand{\thefigure}{{\bf S\arabic{figure}}}
\renewcommand{\theequation}{S\arabic{equation}}%
\newcommand{\tbd}[1]{\textcolor{red}{#1}}

\newpage
\begin{center}
 \Large
{\bf Supplementary Information }
\normalsize
\end{center}
\section{Emergence of Duffing nonlinearity from cavity optomechanics}
\subsection{Brief review of Duffing nonlinearity}
\label{sec: duffing_review}
\begin{figure*}[h!]
    \centering
    \includegraphics[width=0.8\linewidth]{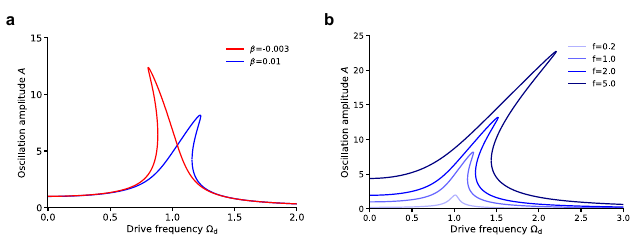}
    \caption{{\bf Duffing nonlinearity.} {\bf a.} Frequency response curve of a Duffing oscillator. Red: softening Duffing response ($\beta<0$). Blue: hardening Duffing response ($\beta>0$). The drive strength is fixed at $f=1$. {\bf b.} Frequency response curves for varying drive strengths $f$ (hardening $\beta=0.01$). The curves in a and b are computed from Eq.~\ref{eq:duffing_A} with $\Omega_\text{m}=1$ and $\Gamma=0.1$.}
    \label{fig:SI_theory_duffing}
\end{figure*}

In this paper, we use an optomechanically-induced Duffing nonlinearity to achieve programmable computation. Before discussing the Duffing nonlinearity in cavity optomechanical systems, we here briefly introduce the main properties of this nonlinearity. The Duffing nonlinearity is a cubic nonlinearity that is described by the equation of motion
\begin{equation}
    \ddot{x} = -\Omega_\text{m}^2 x - \beta x^3 - \Gamma \dot{x} + f\cos\Omega_\text{d} t,
    \label{eq:EOM_duffing_basic}
\end{equation}
where $\Omega_\text{m}$ is the natural frequency, $\beta$ the Duffing nonlinearity coefficient, $\Gamma$ the damping rate, $f$ the external drive strength, and $\Omega_\text{d}$ the external drive frequency.
Neglecting higher frequency terms, the steady-state solution can be written as $x(t)=A\cos(\Omega_\text{d} t+\phi)$, where $A$ and $\phi$ are the amplitude and phase of the response.
The amplitude $A$ satisfies the following form, which is well known as the sharkfin shape\cite{schmid2016fundamentals} (see Fig.~\ref{fig:SI_theory_duffing} a):
\begin{equation}
    A^2 = \frac{f^2}{(-\Omega_\text{d}^2+\Omega^2+\frac{3}{4}\beta A^2)^2+(\Gamma\Omega_\text{d})^2}.
    \label{eq:duffing_A}
\end{equation}

When the Duffing nonlinearity coefficient is positive ($\beta>0$), the system exhibits hardening behavior, and the resonance peak shifts towards higher frequencies as the oscillation amplitude increases. When $\beta<0$, the system shows softening, as shown in Fig.~\ref{fig:SI_theory_duffing} a. As the driving strength $f$ increases, this `bending' of the resonant response spectrum becomes more pronounced (Fig.~\ref{fig:SI_theory_duffing} b). 

\subsection{Duffing nonlinearity induced by the cavity optomechanical interaction}
\label{sec:theory_singlemode}
\subsubsection{Observable displacement $x$: optical change to the mechanical equation of motion}
\label{sec:theory_single_classical}
In the following we provide the theoretical framework that describes the emergence of mechanical nonlinearities through backaction in a cavity optomechanical system. We start from the standard cavity optomechanical Hamiltonian in the rotating frame of the optical field~\cite{aspelmeyer2014cavity}
 \begin{align}
    H=-\hbar(\Delta_0+Gx)aa^*+ \frac{p^2}{2m} + \frac{1}{2}m\Omega_\text{m}^2 x^2, 
    \label{eq:hamiltonian classical}
\end{align}
where $a$ is the cavity field amplitude, $\Delta_0$ is the equilibrium detuning $\Delta_0=\omega_\text{L}-\omega_\text{cav}$, $\omega_\text{L}$ is the external laser frequency, $\omega_\text{cav}$ is the resonance frequency of the optical cavity, $x$ and $p$ are the mechanical displacement and momentum, $m$ is the effective mass of the resonator, $\Omega_\text{m}$ is its mechanical resonance frequency, $G$ is the optomechanical coupling strength (linear cavity frequency shift per unit displacement) and $\hbar$ is the reduced Planck constant. Here the cavity field amplitude $a$ is normalized such that $aa^*$ corresponds to the number of photons inside the cavity (i.e., $a$ corresponds to the annihilation operator in a quantum formalism). This normalization introduces the reduced Planck constant $\hbar$ into the Hamiltonian. Heisenberg equations of motion for the optical and mechanical degrees of freedom can be obtained from this Hamiltonian upon introducing damping~\cite{aspelmeyer2014cavity}. The evolution of the optical field $a$ of an optomechanical cavity driven by an external laser with frequency $\omega_L$, coupling in through a port carrying field $a_\text{in}$ with rate $\kappa_\text{in}$, is given by
\begin{equation}
    \dot{a}=\left(i(\Delta_0+Gx)-\frac{\kappa}{2}\right){a}+\sqrt{\kappa_\text{in}}a_\text{in},
\end{equation}
where $\kappa$ is the total cavity energy decay rate. The mechanical equation of motion (EOM) with mechanical damping rate $\Gamma$ reads
\begin{equation}
    \ddot{x} = -\Omega_\text{m}^2 x-\Gamma\dot{x}+\frac{\hbar G |a|^2}{m},
    \label{eq:EOM_x_orig}
\end{equation}
where the last term represents the radiation pressure force.

We consider systems in the unresolved sideband regime (bad cavity limit) $\kappa\gg\Omega_\text{m}$, meaning that the cavity field instantaneously follows a mechanical displacement. For each mechanical position $x$, we thus assume that the optical field is in its steady state
\begin{equation}
    a_\text{ss}(x)=\frac{\sqrt{\kappa_\text{in}}a_\text{in}}{\kappa/2-i(\Delta_0+Gx)}.
    \label{eq:a_ss}
\end{equation}
Therefore, the cavity photon number $n_\text{c}$ is given by
\begin{equation}
    n_\text{c}=|a_\text{ss}|^2 = n_\text{max}\frac{1}{1+(2(\Delta_0+Gx)/\kappa)^2},
    \label{eq:n_c}
\end{equation}
where $n_\text{max}=4|a_\text{in}|^2\kappa_\text{in}/\kappa^2$ is the maximum number of photons that the drive field can excite in the cavity.
By substituting Eq.~\ref{eq:n_c} into Eq.~\ref{eq:EOM_x_orig}, we obtain the decoupled mechanical EOM
\begin{equation}
    \ddot{x} = -\Omega_\text{m}^2x-\Gamma\dot{x}+\frac{\hbar G n_\text{max}}{m}\frac{1}{1+(2(\Delta_0+Gx)/\kappa)^2}.
\end{equation}

Expanding the last term linearly in $x$ provides an effective shift of the mechanical frequency known as the optical spring effect. Importantly, however, the radiation pressure force is a nonlinear function in terms of the mechanical displacement $x$. We define $f=\bar G|a|^2/m$ to be the force divided by mass $m$. We perform a Taylor expansion in terms of $x$ around $x=0$ to obtain
\begin{align}
    &f(x) = \frac{\hbar G n_\text{max}}{m}\frac{1}{1+(2(\Delta_0+Gx)/\kappa)^2}\nonumber\\
    &f(x)=f(0)+f'(0)x+\frac{f''(0)}{2!}x^2+\frac{f'''(0)}{3!}x^3+\mathcal{O}(x^4)\nonumber\\
    &f(0)=\frac{\hbar G n_\text{max}}{m}\frac{\kappa^2}{4\Delta_0^2+\kappa^2}\nonumber\\
    &f'(0)=-\frac{\hbar G n_\text{max}}{m}\frac{8G\kappa^2\Delta_0}{(4 \Delta _0^2+\kappa^2) ^2}\\\
    &f''(0)=-\frac{\hbar G n_\text{max}}{m}\frac{8G^2\kappa^2(\kappa^2-12\Delta_0^2)}{(4\Delta_0^2+\kappa^2)^3}\nonumber\\
    &f'''(0)=\frac{\hbar G n_\text{max}}{m}\frac{384G^3\kappa^2\Delta_0(\kappa^2-4\Delta_0^2)}{(4\Delta_0^2+\kappa^2)^4}.
\end{align}
Neglecting fourth-order terms and higher, we rewrite the EOM as
\begin{align}
    &\ddot{x} = -\alpha x - \mu x^2 - \beta x^3 -\Gamma\dot{x} + f,\nonumber\\
    &\alpha = \Omega_\text{m}^2+\frac{\hbar G n_\text{max}}{m}\frac{G}{\kappa}\frac{4u_0}{(1+u_0^2)^2},\nonumber\\
    & \mu = \frac{\hbar G n_\text{max}}{m}\frac{G^2}{\kappa^2}\frac{4(1-3u_0^2)}{(1+u_0^2)^3},\nonumber\\
    & \beta = \frac{\hbar G n_\text{max}}{m}\frac{G^3}{\kappa^3}\frac{32u_0(u_0^2-1)}{(1+u_0^2)^4}, \nonumber\\
    & f = \frac{\hbar G n_\text{max}}{m}\frac{1}{1+u_0^2},
\end{align}
where $u_0$ is the normalized optical detuning defined as $u_0=2\Delta_0/\kappa$.
The effective mechanical resonance frequency is $\Omega=\Omega_\text{m}+\delta\Omega$, with an optically induced shift
\begin{equation}
\delta\Omega\approx\frac{\hbar G^2 n_\mathbf{max}}{m\kappa\Omega_\mathbf{m}}\frac{2u_0}{(1+u_0^2)^2} = \frac{g_0^2n_\mathbf{max}}{\kappa}\frac{4u_0}{(1+u_0^2)^2}.
\end{equation}

The optomechanically induced Duffing nonlinearity is parametrized through the coefficient $\beta$. Clearly, $\beta$ can be tuned by changing the laser power, which determines $n_\text{max}$, or by adjusting the laser frequency $\omega_\text{L}$, which modifies the normalized detuning 
$u_0$ to change the magnitude as well as the sign of the nonlinearity.
 
\subsubsection{Complex displacement $b$ and rotating wave approximation (RWA)}
In the remainder of this discussion, we adopt the complex displacement $b$ in place of the real mechanical displacement $x$, since this formulation is more convenient for applying the rotating wave approximation (RWA) and for simplifying the analytical derivations. We treat the mechanical mode in a manner analogous to the optical field, introducing a normalized complex amplitude $b$ such that $bb^*=|b|^2$ represents the number of phonons in the mechanical resonator. The real mechanical displacement $x$ is related to $b$ through the zero-point fluctuation amplitude $x_\text{zpf}$
\begin{align}
    & x=x_\text{zpf}(b+b^*),\\
    & x_\text{zpf}=\sqrt{\frac{\hbar}{2m\Omega_\text{m}}}.
\end{align}
Under this convention, we introduce the vacuum optomechanical coupling strength
\begin{equation}
    g_0 = G x_\text{zpf}.
\end{equation}
With these definitions, the optomechanical Hamiltonian can be conveniently rewritten in terms of the complex mechanical displacement $b$ and $b^*$ in the rotating frame of the optical field
\begin{equation}    
H=-\hbar\Delta_0 a^*{a}+\hbar\Omega_\text{m} b^*{b}-\hbar g_0(b+{b}^*){a}^*{a},
\end{equation}

After applying the rotating-wave approximation (RWA), the dynamics is governed by the equations of motion
\begin{align}
    &\dot{a}=(i(\Delta_0 + g_0(b+b^*))-\frac{\kappa}{2}){a}+\sqrt{\kappa_\text{in}}a_\text{in},\nonumber\\
    &\dot{{b}}=-i\Omega_\text{m} b-\frac{\Gamma}{2}b + ig_0{a}^*{a}, \nonumber\\
    &\dot{b}^*=i\Omega_\text{m} b^* -\frac{\Gamma}{2}b^*-ig_0 a^*a.\label{eq: EOM_b_RWA}
\end{align}

Analogous to the treatment above, in the bad-cavity limit we obtain a decoupled mechanical equation
\begin{align}
    \dot{b}=-i\Omega_\text{m} b-\frac{\Gamma}{2}b +ig_0 n_\text{max}\frac{1}{1+(2(\Delta_0+g_0(b+b^*))/\kappa)^2}.
\end{align}
The last term represents the radiation pressure force, which we again should recognize to be a nonlinear function of $b + b^*$. Because $n_{\text{max}}(t)$ could vary in time, a general rotating-wave approximation (RWA) cannot be applied to drop terms involving specific combinations of $b$ and/or $b^*$. Therefore, we retain both operators $b$ and $b^*$, and apply a Taylor expansion to find
\begin{align}
    \dot{b}=&-i\Omega_\text{m} b-\frac{\Gamma}{2}b+ig_0 n_\text{max}\frac{1}{1+u_0^2}\nonumber\\
    &-\frac{4i g_0^2 n_\text{max}}{\kappa}\frac{u_0}{(1+u_0^2)^2}(b+b^*)\nonumber\\
    &-\frac{4i g_0^3 n_\text{max}}{\kappa^2}\frac{1-3u_0^2}{(1+u_0^2)^3}(b+b^*)^2\nonumber\\
    &-\frac{32i g_0^4 n_\text{max}}{\kappa^3}\frac{u_0 (u_0^2-1)}{(1+u_0^2)^4}(b+b^*)^3+\mathcal{O}^4.
    \label{eq:single_mode_taylor}
\end{align}
Again, the Duffing nonlinearity can be recognized in the third-order term.

\subsubsection{Multiple lasers}
In the reported experiments, we use up to four independent lasers. When multiple lasers are present, the general equation of motion becomes
\begin{align}
    \dot{b}&=-i\Omega_\text{m} b-\frac{\Gamma}{2}b-i\alpha(b+b^*)-i\mu(b+b^*)^2-i\beta(b+b^*)^3+i f,\nonumber\\
    &=-i\Omega_\text{m} b-\frac{\Gamma}{2}b-\frac{4i g_0^2}{\kappa}\alpha'(b+b^*)-\frac{4i g_0^3}{\kappa^2}\mu'(b+b^*)^2-\frac{32i g_0^4}{\kappa^3}\beta'(b+b^*)^3+ig_0f',\label{eq:singel_mode_multilaser_1}\\
    \alpha'&=\sum_l n_l(t)\frac{u_l}{(1+u_l^2)^2},\nonumber\\
    \mu'&=\sum_l n_l(t)\frac{1-3u_l^2}{(1+u_l^2)^3},\nonumber\\
    \beta'&=\sum_l n_l(t)\frac{u_l (u_l^2-1)}{(1+u_l^2)^4},\nonumber\\
    f'&=\sum_l n_l(t)\frac{1}{1+u_l^2},
    \label{eq:singel_mode_multilaser_2}
\end{align}
where $n_l$ is maximum photon number $n_\text{max}$ in the cavity that can be excited by the laser $l$, and $u_l$ is the normalized optical detuning $u_0$ of laser $l$. It is written here as a function of time $t$ to accomodate possible temporal intensity modulation.

\subsubsection{Selecting optical frequency}

\label{sec:theory_selecting optical frequency}
In the single-mode experiments we employ three lasers with distinct roles: (i) a detection laser, used solely for readout; (ii) a function laser, used to engineer the optomechanical Duffing nonlinearity; and (iii) a drive laser with time-modulated intensity, used to introduce digital signals into the system. The detection laser is placed far off resonance so that all its induced coefficients are negligible. The function laser is tuned to $u_\text{f}=\pm\sqrt{1-\frac{2}{5}\sqrt{5}}$ to maximize the Duffing coefficient $\beta'$. In this paper, we choose softening $u_\text{f}=\sqrt{1-\frac{2}{5}\sqrt{5}}$. The signal laser is set to zero detuning $u_\text{s}=0$ to avoid interference with the function laser and maximize the driving strength.

\subsubsection{Oscillation amplitude under selected parameters}
Because the intensity modulator is not linear over its full drive voltage range, we apply a DC bias to operate in the linear region and suppress modulator-induced nonlinearities, see Methods and Fig.~\ref{fig:SI_modulator_linear} for details. Consequently, for the signal (drive) laser the maximum cavity photon number varies in time as $n_\text{d}(t) = n_\text{d}^0 +  n_\text{d}'\cos (\Omega_\text{d}t+\phi_\text{d})$. The maximum cavity photon number $n_\text{f}$ induced by the function laser remains constant. 
We obtain the coefficients
\begin{align}
    &\alpha' = \alpha'_\text{f} + \alpha'_\text{d}=0.266 n_\text{f},\nonumber\\
    &\mu' = \mu'_\text{f} + \mu'_\text{d}=0.506n_\text{f}+n_\text{d}^0+n'_\text{d}\cos(\Omega_\text{d}t+\phi_\text{d}),\nonumber\\
    &\beta'=\beta'_\text{f} + \beta'_\text{d} = -0.195n_\text{f},\nonumber\\
    & f' = f'_\mathbf{f} + f'_\mathbf{d} = 0.905 n_\mathbf{f} + n_\text{d}^0 +  n_\text{d}'\cos (\Omega_\text{d}t+\phi_\text{d})
    \label{eq:single_mode_real_model_coefficient}
\end{align}
We assume that the solution is 
\begin{equation}
    b = \frac{A}{2}e^{-i\Omega_\text{d}t},
    \label{eq:single_mode_assume_A}
\end{equation}
where, without loss of generality, we set $A$ real by redefining the phase origin $\phi_\text{d}$.
By substituting Eqs.~\ref{eq:single_mode_real_model_coefficient} and \ref{eq:single_mode_assume_A} to Eqs.~\ref{eq:singel_mode_multilaser_1} and \ref{eq:singel_mode_multilaser_2} and applying the rotating-wave approximation (RWA), we obtain
\begin{align}
   \left(\Omega_\text{d}-\Omega_\text{m}+i\frac{\Gamma}{2}-\frac{1.064g_0^2}{\kappa}n_\text{f}\right)A - \frac{g_0^3}{\kappa^2}n'_\text{d}\left(e^{i\phi_\text{d}}+2e^{-i\phi_\text{d}}\right)A^2+\frac{4.68g_0^4}{\kappa^3}n_\text{f}A^3+g_0n'_\text{d}e^{-i\phi_\text{d}}=0.
   \label{eq:single_model_solution_0}
\end{align}
We rewrite the Eq.~\ref{eq:single_model_solution_0} as
\begin{align}
    &\left(\delta\Omega_\text{d}+i\frac{\Gamma}{2}\right)A-\mu_\text{eff}\left(3\cos\phi_\text{d}-i\sin\phi_\text{d}\right)A^2-\beta_\text{eff}A^3+f_\text{eff}\left(\cos\phi_\text{d}-i\sin\phi_\text{d}\right)=0\nonumber\\
    &\delta\Omega_\text{d}=\Omega_\text{d}-\left(\Omega_m+\frac{1.064g_0^2}{\kappa}n_\text{f}\right)\nonumber\\
    &\mu_\text{eff}=\frac{g_0^3}{\kappa^2}n'_\text{d}=\frac{g_0^2}{\kappa^2}f_\text{eff}\nonumber\\
    &\beta_\text{eff}=-\frac{4.68g_0^4}{\kappa^3}n_\text{f}\nonumber\\
    &f_\text{eff}=g_0n'_\text{d}.
    \label{eq:single_model_solution_1}
\end{align}

To solve Eq.~\ref{eq:single_model_solution_1}, we set its real and imaginary parts to zero:
\begin{align}
    &\delta\Omega_\text{d}A-3\mu_\text{eff}\cos\phi_\text{d}A^2-\beta_\text{eff}A^3+f_\text{eff}\cos\phi_\text{d}=0,\nonumber\\
    &-\frac{\Gamma}{2}A-\mu_\text{eff}\sin\phi_\text{d}A^2+f_\text{eff}\sin\phi_\text{d}=0.
\end{align}
The final implicit expression for the oscillation amplitude $A$ is
\begin{align}
    \frac{f_\text{eff}^2}{A^2}=\frac{\Gamma^2}{4\left(1-g_0^2A^2/\kappa^2\right)^2}+\frac{\left(\delta\Omega_\text{d}-\beta_\text{eff}A^2\right)^2}{\left(1-3g_0^2A^2/\kappa^2\right)^2}.
    \label{eq:A_quadratic}
\end{align}
Eq.~\ref{eq:A_quadratic} differs from the standard Duffing response in Eq.~\ref{eq:duffing_A} for three reasons: 

(i) the derivation is carried out in terms of the complex displacement $b$, which leads to a slightly different RWA. This difference is negligible for large quality factors $Q\gg1$ (here, $Q\sim 10^3$) and for drive frequency $\Omega_\text{d}$ near the natural frequency $\Omega_\text{m}$;

(ii) the control laser not only induces Duffing nonlinearity but also shifts the effective eigenfrequency through the optical spring effect. This shift is accounted for by the experimental calibration described in Methods;

(iii) the modulation of the drive laser, together with optomechanical coupling, makes the second quadratic term non-negligible under the RWA.

\subsection{Coupling different eigenmodes using optomechanical backaction}
In Sec.~\ref{sec:theory_singlemode}, we considered only one mechanical mode. To implement the possibility for cascading logic gates, we introduce laser-induced coupling between different eigenmodes in a multimode resonator. In this section, we show how optomechanical interactions enable this coupling. We follow the linear theory introduced in Ref. \cite{delpino2022, mathew2020synthetic}, and then continue to discuss the influence of the nonlinearity of the spring effect.

\subsubsection{General EOM}
The EOM for eigenmode $i$, characterized by resonance frequency $\Omega_{\text{m},i}$, damping rate $\Gamma_i$ and optomechanical coupling strength $g_i$, is
\begin{align}
    \dot{b}_i=-i\Omega_{\text{m},i}b_i-\frac{\Gamma_i}{2}b_i+i g_i\sum_{l} n_{l}(t)\frac{1}{1+\left(u_{l}+\sum_{j}^N 2g_j \left(b_j+b_j^*\right)/\kappa\right)^2 },
    \label{eq:EOM_multimode}
\end{align}
where $j\in\{{1,\dots,N}\}$ indexes the mechanical eigenmodes and $l$ indexes different lasers, each characterized by a normalized optical detuning \(u_l = 2\Delta_{0,l}/\kappa\) and a maximum photon number in the cavity $n_{\text{max},l}$.

We define a function $h_l$ as
\begin{equation}
    h_{l}=\frac{1}{1+\left(u_l+\sum_{j}^N 2g_j\left(b_j+b_j^*\right)/\kappa\right)^2 }.
\end{equation}
Defining $z_j=b_j+b_j^*$, we take the Taylor expansion of $h_l(\mathbf{z})$ around $\mathbf{z}=(z_1,z_2,...,z_N)=(0,0,...,0)=\mathbf{0}$ to find
\begin{align}
    h_l(\mathbf{z})=&h_l(\mathbf{z}=\mathbf{0})+\sum_{j=1}^N\frac{\partial h_l}{\partial z_j} \bigg|_{\mathbf{z}=\mathbf{0}}z_j+\frac{1}{2!}\sum_{j=1}^N\sum_{k=1}^N\frac{\partial^2 h_l}{\partial z_j\partial z_k}\bigg|_{\mathbf{z}=\mathbf{0}}z_j z_k\nonumber\\
    &+\frac{1}{3!}\sum_{j=1}^N\sum_{k=1}^N\sum_{m=1}^N\frac{\partial^3 h_l}{\partial z_j\partial z_k\partial z_m}\bigg|_{\mathbf{z}=\mathbf{0}}z_j z_k z_m+\mathcal{O}(z_j^4)\nonumber\\
    \frac{\partial h_l}{\partial z_j} \bigg|_{\mathbf{z}=\mathbf{0}}&=-\frac{4 g_j}{\kappa}\frac{u_l}{(1+u_l^2)^2}\nonumber\\
    \frac{\partial^2 h_l}{\partial z_j\partial z_k} \bigg|_{\mathbf{z}=\mathbf{0}}&=-\frac{8 g_j g_k}{\kappa^2}\frac{1-3u_l^2}{(1+u_l^2)^3}\nonumber\\
    \frac{\partial^3 h_l}{\partial z_j\partial z_k\partial z_m}\bigg|_{\mathbf{z}=\mathbf{0}}&=-\frac{192 g_j g_k g_m}{\kappa^3}\frac{u_l (u_l^2-1)}{(1+u_l^2)^4}.
    \label{eq:taylor_multimode}
\end{align}
By substituting Eq.~\ref{eq:taylor_multimode} into Eq.~\ref{eq:EOM_multimode}, we obtain
\begin{align}
    \dot{b}_i=&-i\Omega_{\text{m},i}b_i-\frac{\Gamma_i}{2}b_i+i g_i\sum_l n_l\frac{1}{1+u_l^2}\nonumber\\
    &-\frac{4ig_i}{\kappa}\sum_ln_l\frac{u_l}{(1+u_l^2)^2}\sum_{j=1}^N g_j (b_j+b_j^*)\nonumber\\
    &-\frac{4ig_i }{\kappa^2}\sum_l n_l\frac{1-3u_0^2}{(1+u_0^2)^3}\sum_{j=1}^N\sum_{k=1}^N g_j g_k(b_j+b_j^*)(b_k+b_k^*)\nonumber\\
    &-\frac{32i g_i}{\kappa^3}\sum_l n_l\frac{u_l(u_l^2-1)}{(1+u_l^2)^4}\sum_{j=1}^N\sum_{k=1}^N\sum_{m=1}^N g_j g_k g_m (b_j+b_j^*)(b_k+b_k^*)(b_m+b_m^*).
    \label{eq:EOM_multimode_general}
\end{align}

We note that, since all eigenmodes are present, this is the standard EOM. In Sec.~\ref{sec:theory_singlemode} we only considered a single mechanical mode. Indeed, in the absence of a coupling laser whose intensity is modulated near the frequency difference between coupled eigenmodes, the other terms with indices $j\neq i$, $k \neq i$ and $m\neq i$ are neglected during the RWA.

\subsubsection{Coupled resonators in the Rotating Wave Approximation}
\label{sec:twomode_RWA}
We now the discuss the introduction of a coupling laser, for the specific case of coupling two modes through a difference frequency modulation. We retain the terms associated with the other lasers that we found above by neglecting the remaining cross-mode terms in the RWA in absence of the coupling laser, and isolate here the contribution of the coupling laser. We set the normalized optical detuning of the coupling laser to $u_{\text{c}}=\pm 1/\sqrt{3}$, which maximizes the spring effect and therefore the coupling. To choose the detuning sign, we note that the function laser is chosen to produce a softening Duffing response ($u_{\text{f}}>0$). Although the coupling laser intensity is modulated by an AC signal, technical constraints discussed in the Methods and Fig.~\ref{fig:SI_modulator_linear} require a finite DC component as well. This DC component also adds Duffing nonlinearity; to avoid cancellation between the Duffing contributions of the two lasers, we therefore choose $u_{\mathrm{c}}=+ 1/\sqrt{3}>0$. At this detuning, the quadratic coefficient contributed by the coupling laser vanishes ($\mu'_{\text{c}}=0$). For simplicity, we also neglect all cubic coupling terms with indices $j\neq i$, $k\neq i$, and $m\neq i$ in Eq.~\ref{eq:EOM_multimode_general}. This simplification ignores a possible nonlinear coupling (cross-Duffing effect). We acknowledge such nonlinear coupling may well play a role in realistic systems. However, its effect on the mode with the largest Duffing nonlinearity is generally smaller than the effect of the self-Duffing term, since the nonlinear terms in eq.~\ref{eq:EOM_multimode_general} with $g_i^3g_j/\kappa^3$, $g_i^2g_j^2/\kappa^3$, or $g_ig_j^3/\kappa^3$ scaling will be samller than the $g_i^4/\kappa^3$ term if $g_j<g_i$. We write the new EOM where modes $i$ and $j$ are considered as
\begin{align}
    \dot{b}_i=&-i\Omega_{\text{m},i}b_i-\frac{\Gamma_i}{2}b_i+\text{contribution from other lasers}\\
    &+i g_i n_\text{c}(t)\frac{1}{1+u_\text{c}^2}\\
    &-\frac{4ig_i^2 n_\text{c}(t)}{\kappa}\frac{u_\text{c}}{(1+u_\text{c}^2)^2}(b_i+b_i^*)\label{eq:EOM_twomodes_s1}\\
    &-\frac{4ig_ig_j n_\text{c}(t)}{\kappa}\frac{u_\text{c}}{(1+u_\text{c}^2)^2}(b_j+b_j^*)\label{eq:EOM_twomodes_s2}\\
    &-\frac{32i g_i^4n_\text{c}(t)}{\kappa^3}\frac{u_\text{c} (u_\text{c}^2-1)}{(1+u_\text{c}^2)^4}(b_i+b_i^*)^3\label{eq:EOM_twomodes_s3}.
\end{align}
Here, the last term corresponds to the Duffing nonlinearity caused by the coupling lasers. Its DC component can be lumped with that of the function laser. 

We begin with a simple case in which only two mechanical modes $i= 1$ and $j=3$ are considered. To couple these two modes, we modulate the intensity of the coupling laser at a frequency close to their frequency difference, $\Omega_{\text{c}}\approx \Omega_{\text{m},3}-\Omega_{\text{m},1}$. The maximum photon number induced by the coupling laser is $n_{\text{c}}(t)=n_{\text{c}}^{0}+n_{\text{c}}'\cos(\Omega_{\text{c}} t+\phi_{\text{c}})$. Under the RWA, the coupling laser contributes three terms. Its DC component, $n_\text{c}^0$, leads to an additional optical spring effect and a Duffing nonlinearity through Eq.~\ref{eq:EOM_twomodes_s1} and Eq.~\ref{eq:EOM_twomodes_s3}. The AC component, $n_{\text{c}}'\cos(\Omega_{\text{c}} t+\phi_{\text{c}})$, brings the coupling between the two modes.

Now assume mode 1 is driven by an external laser whose intensity is modulated at $\Omega_\text{d}$ with phase $\phi_\text{d}$. We couple modes 1 and 3 using the coupling laser. We assume the solutions for $b_1$ and $b_3$ are:
\begin{equation}
    b_1=\frac{A_1}{2}e^{-i\Omega_\text{d}t}, \hspace{0.5cm} b_3=\frac{A_3}{2}e^{-i(\Omega_\text{d}+\Omega_\text{c})t},
\end{equation}
where, without loss of generality, we assume $A_1$ is a real number and $A_3$ is a complex number.

Applying the RWA, we obtain the coupled equations for $A_1$ and $A_2$:
\begin{align}
    &\left(\delta\Omega_1+i\frac{\Gamma_1}{2}\right)A_1-\alpha_\text{eff}e^{i\phi_\text{c}}A_3-\mu_\text{eff}(3\cos\phi_\text{d}-i\sin\phi_\text{d})A_1^2-\beta_{\text{eff},1}A_1^3+f_\text{eff}e^{-i\phi_\text{d}}=0\nonumber\\
    &\left(\delta\Omega_3+i\frac{\Gamma_3}{2}\right)A_3-\alpha_\text{eff}e^{-i\phi_\text{c}}A_1-\beta_{\text{eff},3}A_3^2A_3^*=0\label{eq:A_twomode_coupled}\\
    &\delta\Omega_1=\Omega_\text{d}-\left(\Omega_{\text{m},1}+\frac{1.064g_1^2}{\kappa}n_\text{f}+\frac{1.299g_1^2}{\kappa}n_\text{c}^0\right)\nonumber\\
     &\delta\Omega_3=\Omega_\text{d}+\Omega_\text{c}-\left(\Omega_{\text{m},3}+\frac{1.064g_3^2}{\kappa}n_\text{f}+\frac{1.299g_3^2}{\kappa}n_\text{c}^0\right)\nonumber\\
    &\alpha_{\text{eff}} = \frac{0.650g_1g_3}{\kappa}n'_\text{c}\nonumber\\
    &\mu_{\text{eff}}=\frac{g_i^3}{\kappa^2}n'_\text{d}\nonumber\\
    &\beta_{\text{eff},i}=-\left(\frac{4.68g_i^4}{\kappa^3}n_\text{f}+\frac{2.923g_i^4}{\kappa^3}n_c^0\right)\nonumber\\
    &f_\text{eff}=g_1n'_\text{d}
\end{align}

Eq.~\ref{eq:A_twomode_coupled} is too complicated to solve analytically; therefore, we simplify it by assuming mode 1 is linear and mode 3 is nonlinear. This is a simplified description of the experimental case since generally $|g_j|<|g_i|$. We thus obtain the simplified version:
\begin{align}
    &\left(\delta\Omega_1+i\frac{\Gamma_1}{2}\right)A_1-\alpha_\text{eff}e^{i\phi_\text{c}}A_3+f_\text{eff}e^{-i\phi_\text{d}}=0\nonumber\\
    &\left(\delta\Omega_3+i\frac{\Gamma_3}{2}\right)A_3-\alpha_\text{eff}e^{-i\phi_\text{c}}A_1-\beta_{\text{eff},3}A_3^2A_3^*=0
    \label{eq:A_twomode_coupled_simplfied}
\end{align}

From the first equation, we obtain the expression for $A_1$
\begin{align}
    A_1=\frac{\alpha_\text{eff}e^{i\phi_\text{c}}A_3-f_\text{eff}e^{-i\phi_\text{d}}}{\delta\Omega_1+i\Gamma_1/2}.
    \label{eq:A1_twomode}
\end{align}
By substituting Eq.~\ref{eq:A1_twomode} into Eq.~\ref{eq:A_twomode_coupled_simplfied}, we obtain the implicit expression for $A_3$:
\begin{align}
    \left(\delta\Omega_3+i\frac{\Gamma_3}{2}\right) A_3-\beta_{\text{eff},3} A_3^2 A_3^*-\alpha_\text{eff}e^{-i\phi_\text{c}}\frac{\alpha_\text{eff}e^{i\phi_\text{c}}A_3-f_\text{eff}e^{-i\phi_\text{d}}}{\delta\Omega_1+i\Gamma_1/2}=0.
\end{align}
We further simplify the expression as
\begin{equation}
     \left(\delta\Omega_3+i\frac{\Gamma_3}{2}-\frac{\alpha_\text{eff}^2}{\delta\Omega_1+i\Gamma_1/2}-\beta_{\text{eff},3} |A_3|^2\right) A_3  + \frac{\alpha_\text{eff}f_\text{eff}e^{-i(\phi_\text{c}+\phi_\text{d})}}{\delta\Omega_1+i\Gamma_1/2}=0.
\end{equation}
After solving the expression above to obtain the value of $A_3$, the solution for $A_1$ can be calculated using Eq.~\ref{eq:A1_twomode}.

Finally, we include for completeness the simplified description of a three-mode system under the same assumptions, taking mode 3 as a nonlinear mode coupled to the linear resonators 1 and 2. Starting from the EOMs
\begin{align}
    \dot{b}_1=&-i\Omega_{\text{eff},1} b_1 -\frac{\Gamma_1}{2}b_1-i\alpha_{13}\cos(\Omega_{\text{c},13} t +\phi_{13})(b_3+b_3^*)+if_1\cos(\omega_{1} t+\phi_{\text{d},1})\nonumber\\
     \dot{b}_2=&-i\Omega_{\text{eff},2} b_2 -\frac{\Gamma_2}{2}b_2-i\alpha_{23}\cos(\Omega_{\text{c},23}+\phi_{23})(b_3+b_3^*)+if_2\cos(\omega_{2} t+\phi_{\text{d},2})\nonumber\\
    \dot{b}_3=&-i\Omega_{\text{eff},3} b_3 -\frac{\Gamma_3}{2}b_3-i\alpha_{13}\cos(\Omega_{\text{c},13} t +\phi_{13}) (b_1+b_1^*)-i\alpha_{23}\cos(\Omega_{\text{c},23} t +\phi_{23})(b_2+b_2^*)-i\beta_3(b_3+b_3^*)^3,
\end{align}
where $\Omega_{\text{eff},i}$ is the effective eigenfrequency that includes the optomechanically induced spring effect, we obtain the coupled equations for the oscillation amplitudes $A_i$ using the RWA:
\begin{align}
    &\left(\delta\Omega_1+i\frac{\Gamma_1}{2}\right)A_1-\alpha_{\text{eff},13}e^{i\phi_{\text{c},13}}A_3+f_{\text{eff},1}e^{-i\phi_{\text{d},1}}=0\nonumber\\
    &\left(\delta\Omega_2+i\frac{\Gamma_2}{2}\right)A_2-\alpha_{\text{eff},23}e^{i\phi_{\text{c},23}}A_3+f_{\text{eff},2}e^{-i\phi_{\text{d},2}}=0\nonumber\\
    &\left(\delta\Omega_3+i\frac{\Gamma_3}{2}\right)A_3-\alpha_{\text{eff},13}e^{-i\phi_{\text{c},13}}A_1-\alpha_{\text{eff},23}e^{-i\phi_{\text{c},23}}A_2-\beta_{\text{eff},3}A_3^2A_3^*=0
\end{align}
where 
\begin{align}         &\beta_{\text{eff},3}=3\beta_3/4,\hspace{0.5cm}\alpha_{\text{eff},ij} = \alpha_{ij}/2,\hspace{0.5cm}f_{\text{eff},i} = f_{i}, \nonumber\\
   &\delta\Omega_1= \Omega_{\text{d},1}-\Omega_{\text{eff},1},\hspace{1cm}\delta\Omega_2= \Omega_{\text{d},2}-\Omega_{\text{eff},2},\nonumber\\ &\delta\Omega_3= \Omega_{\text{d},1}+\Omega_{\text{c},13}-\Omega_{\text{eff},3}=\Omega_{\text{d},2}+\Omega_{\text{c},23}-\Omega_{\text{eff},3}.
\end{align}

The implicit expression of $A_3$ is
\begin{align}
    \left(\delta\Omega_3+i\frac{\Gamma_3}{2}-\frac{\alpha_{\text{eff},13}^2}{\delta\Omega_1+i\Gamma_1/2}-\frac{\alpha_{\text{eff},23}^2}{\delta\Omega_2+i\Gamma_2/2}-\beta_{\text{eff},3} |A_3|^2\right) A_3+\frac{\alpha_{\text{eff},13}f_{\text{eff},1}e^{-(\phi_{\text{c},13}+\phi_{\text{d},1})}}{\delta\Omega_1+i\Gamma_1/2}+\frac{\alpha_{\text{eff},23}f_{\text{eff},2}e^{-(\phi_{\text{c},23}+\phi_{\text{d},2})}}{\delta\Omega_2+i\Gamma_2/2}=0.
\end{align}
The oscillation amplitudes $A_1$ and $A_2$ are given by
\begin{align}
    &A_1=\frac{\alpha_{\text{eff},13}e^{i\phi_{\text{c},13}}A_3-f_{\text{eff},1}e^{-i\phi_{\text{d},1}}}{\delta\Omega_1+i\Gamma_1/2},\nonumber\\
    &A_2=\frac{\alpha_{\text{eff},23}e^{i\phi_{\text{c},23}}A_3-f_{\text{eff},2}e^{-i\phi_{\text{d},2}}}{\delta\Omega_2+i\Gamma_2/2}.
\end{align}

\newpage
\section{Computing with Duffing nonlinearity}
\subsection{Logic gates using Duffing nonlinearity} 
\begin{figure}[h!]
    \centering
    \includegraphics[width=0.5\linewidth]{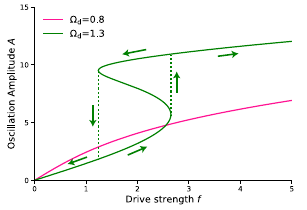}
    \caption{{\bf Hysteresis arising from the Duffing nonlinearity.} The result is computed using Eq.~\ref{eq:duffing_A} with coefficients $\Omega_\text{m}=1$, $\beta=0.01$ and $\Gamma=0.1$.}
    \label{fig:SI_theory_duffing_driveamp}
\end{figure}

In this work, we use the drive strength $f$ in Eq.~\ref{eq:EOM_duffing_basic} as the input signal to construct logic gates. Therefore, we first examine the system’s response as a function of the drive strength $f$ (instead of the drive frequency $\Omega_\text{d}$, which was discussed in Sec.~\ref{sec: duffing_review}). The general behavior is shown in Fig.~\ref{fig:SI_theory_duffing_driveamp}. 
The nonlinear response can lead to multistability, manifested as an ``S-shaped'' response as a function of drive strength (Fig.~\ref{fig:SI_theory_duffing_driveamp}). For the hardening case ($\beta>0$), this occurs when the drive detuning is positive $\delta\Omega_\text{d}=\Omega-\Omega_\text{d}>0$ while for the softening case ($\beta<0$) it occurs when $\delta\Omega_\text{d}<0$. This multistability, along with the associated switching (“jump”) behavior, is utilized to implement the logic gate. The drive strength $f$ serves as the input(s), while the mechanical oscillation amplitude $A$ represents the output. Both the input and output are encoded in amplitude levels, where the lower level corresponds to logical ‘0’ and the higher level corresponds to logical ‘1’.

In experimental drive strength sweeps, the multistability behavior manifests as hysteresis, as illustrated in Fig.~\ref{fig:SI_jump_repeat50} and indicated by the dashed line in Fig.~\ref{fig:SI_theory_duffing_driveamp}. When the drive strength is increased, the system follows a different trajectory compared to when it is decreased. Interestingly, such a hysteretic behavior could be used to realize a sequential logic gate. In this work, however, we focus on combinational logic gates. Therefore, we aim to minimize the hysteresis by carefully choosing the drive frequency $\Omega_\text{d}$, as shown in \fupdate{\EDF A2 c}.

\subsection{Cascading computation constraint}
\label{sec:cascading_constraint}
\subsubsection{Linear Response}
\begin{figure*}[h!]
    \centering
    \includegraphics[width=\linewidth]{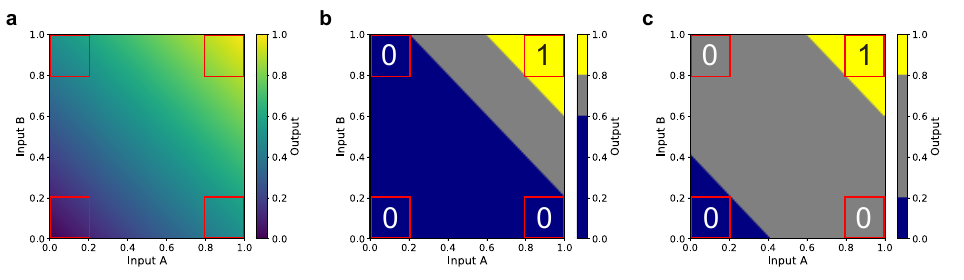}
    \caption{{\bf Cascading computation constraint prevents AND gate realization with linear response.} {\bf a.} Theoretical scan of the system under a linear input–output response. {\bf b.} An AND logic gate can, in principle, be implemented using a linear input–output response when no constraints on input and output ranges are imposed. {\bf c.} An AND logic gate cannot be realized once the cascading requirement is enforced, namely that the output bit accuracy must be better than or equal to the input bit accuracy. The red boxes indicate the defined input accuracy ranges, and the numbers inside the boxes represent the expected values for an AND logic gate. The colorbar represents the output accuracy range, where blue corresponds to bit 0, yellow corresponds to bit 1, and grey denotes outputs that cannot be assigned to either bit value.}
    \label{fig:theory linear logic gate}
\end{figure*}

In the experimental demonstrations of logic gates, we had required the input and output ranges of amplitudes that encode either a 0 or 1 bit to be equal. Here we explain why this is generally a reasonable minimal requirement for any digital computing architecture, and demonstrate that it can be achieved in our system. In fact, if this constraint is relaxed, even a linear response could be used to implement a specific logic gate in a single resonator, as shown in Fig.~\ref{fig:theory linear logic gate} a and b. 
However, this approach does not extend to multi-gate computational networks. This is because the encoding then relies on defining a very precise input range corresponding to bit 0 and bit 1, (indicated by the red boxes in Fig.~\ref{fig:theory linear logic gate} b) while permitting a broader output range (illustrated by the color scale in Fig.~\ref{fig:theory linear logic gate} b). Such outputs, however, are too noisy and cannot reliably provide precise information flow to the next computational unit: the output range of a first gate is incompatible with the input range of the second gate. Moreover, this would preclude the error-correcting nature of digital logic. Consequently, a key requirement for cascading computations is that the accuracy of the output bits must be at least equal to, if not better than, the accuracy of the input bits~\cite{dally2012digital, miller2009device}. 
With this limitation in mind, we observe that a purely linear response cannot be used to realize a logic gate. For instance, implementing an AND gate under these conditions results in an error rate of 50\% (Fig.~\ref{fig:theory linear logic gate} c), even without accounting for any additional noise in the system.

\subsubsection{Cascading constraint}
\label{sec:cascading constraint}
We now introduce the response requirements needed to satisfy the cascading constraint. 
We assume that the response function $f$ is monotonically increasing and satisfies $f(0)=0$. 
Under this assumption, we set the minimum value for both the input and the output encoding of bit~0 to be 0. 
We define the input ranges for bits 0 and 1 as
\begin{equation}
    \mathrm{bit0}_{\mathrm{in}} \in [0, R_0 B], 
    \qquad 
    \mathrm{bit1}_{\mathrm{in}} \in [(1-R_1)B, B],
\end{equation}
where $B$ is the maximum input value.

\begin{figure}
    \centering
    \includegraphics[width=\linewidth]{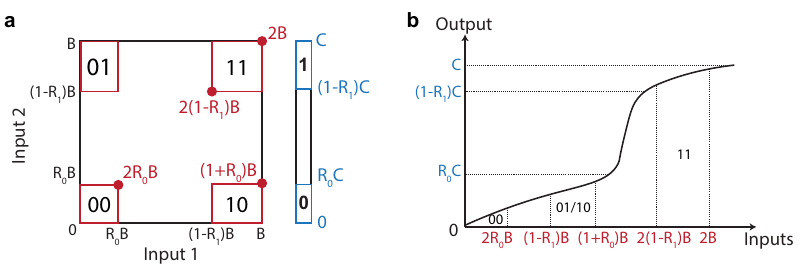}
    \caption{{\bf Cascadable computation constraint. }{\bf a.} Abstract model of a logic gate. The four red boxes indicate the selected input ranges corresponding to inputs 00, 01, 10, and 11. The right panel indicates the output encoding. The labels next to the red dots indicate the sum of the two inputs. {\bf b.} A monotonically increasing response used to implement an AND gate.}
    \label{fig:SI_cascading constraint_v2}
\end{figure}

Because both lasers exert force on the nanobeam simultaneously and with the same phase, the mechanical response is determined by the combined input, i.e., by the sum of the two input signals. For input states 00, 01 or 10, and 11, the corresponding ranges of the summed input are $[0, 2R_0B]$, $[(1-R_1)B, (1+R_0)B]$, and $[2(1-R_1)B, 2B]$, respectively, as shown in Fig.~\ref{fig:SI_cascading constraint_v2}a. 
We further define $C$ as the maximum output value for bit 1. 
Under the cascading constraint, the output encoding must be more accurate than the input encoding.
We take an AND gate as an example and express the cascading constraints as follows, as illustrated in Fig.~\ref{fig:SI_cascading constraint_v2}b.
\begin{align}
    (1+R_0)B &< 2(1-R_1)B, \label{eq:casc1}\\
    f\!\big((1+R_0)B\big) &\leq R_0 C,  \label{eq:casc2}\\
    f\!\big(2(1-R_1)B\big) &\geq (1-R_1) C,  \label{eq:casc3}\\
    f(2B) &\leq C.  \label{eq:casc4}
\end{align}
    
\subsubsection{Cascading constraints on the nonlinear response}
There are four independent parameters that define the logic-gate encoding, namely $B$, $C$, $R_0$, and $R_1$. 
We first seek a qualitative understanding of what type of response function $f$ can satisfy Eqs.~\ref{eq:casc1}-\ref{eq:casc4}. 
Ideally, the output should be as precise as possible. In the absence of noise, this would correspond to a two-plateau (threshold-like) response. 
Although such an idealized threshold is not identical to the Duffing response in Eq.~\ref{eq:EOM_duffing_basic}, it motivates the design goal that the bit-0 and bit-1 regions of $f$ be as flat as possible, separated by a large gap. 
Below we formalize this intuition.

First, Eq.~\ref{eq:casc1} implies
\begin{equation}
    R_0 + 2R_1 < 1 .
    \label{eq:casc5}
\end{equation}
Combining Eqs.~\ref{eq:casc2}-~\ref{eq:casc3} further yields a lower bound on the required ``gap ratio'',
\begin{equation}
    R_{\mathrm{gap}}
    = \frac{f\!\big(2(1-R_1)B\big)}{f\!\big((1+R_0)B\big)}
    > \frac{(1-R_1)C}{R_0C}
    = \frac{1-R_1}{R_0}.
\end{equation}
Using the inequality $R_1 < \tfrac{1}{2}(1-R_0)$ from Eq.~\ref{eq:casc5}, we obtain
\begin{equation}
    R_\mathrm{gap}>\frac{1-R_1}{R_0} > \frac{1}{2}\left(1+\frac{1}{R_0}\right).
\end{equation}
This shows that if $R_0$ is small, the required ratio $R_{\mathrm{gap}}$ becomes very large. 
Conversely, if $R_0$ is large, the constraint forces $R_1$ to be small, which in turn requires $f$ to be very flat in the ``11'' region.

For simplicity and to gain intuition, we consider the symmetric choice $R_0=R_1=R$, which we also adopt in our experiments. 
Eq.~\ref{eq:casc5} then gives
\begin{equation}
    R < \frac{1}{3}.
\end{equation}
In this symmetric case,
\begin{equation}
    R_{\mathrm{gap}} > \frac{1-R}{R} = \frac{1}{R}-1 > 2.
\end{equation}

\subsubsection{Duffing nonlinearity}
Duffing oscillators can exhibit hysteresis. If binary bits are defined within the hysteretic regime, the output depends not only on the current input but also on the previous state of the system. Since our goal is to realize combinatorial logic, where the output should depend solely on the present input and not on the system history, it is essential to ensure that the valid input values lie outside the hysteretic regime.
Consequently, a larger hysteresis range forces us to select smaller values of $R_0$ and $R_1$, which in turn requires flatter output regions for both bit~0 and bit~1. 
Moreover, the hysteresis range cannot be arbitrarily large. If it becomes too wide, the AND gate condition cannot be satisfied.

To see this, start from:
\begin{align}
    \dot{b}=&-i\Omega_\text{m}b-\frac{\Gamma}{2}b -i\alpha(b+b^*)-i\beta(b+b^*)^3+if\cos(\Omega_\text{d} t+\phi_\text{d}),
    \label{eq:duffing_b}
\end{align}
where all coefficients are time independent. We ignore the quadratic term $\mu(b+b^*)^2$ because the quadratic term is neglected during RWA when $\mu$ is constant. 

Taking the ansatz 
\begin{align}
    b = \frac{A}{2}e^{-i\Omega_\text{d} t}
\end{align}
where, without loss of generality, $A$ is chosen real by redefining the phase origin $\phi_\text{d}$, yields
\begin{align}
    &\left[\left(\Omega_\text{d}-\Omega_\text{m}-\alpha-\frac{3}{4}\beta A^2\right)^2+\frac{\Gamma^2}{4}\right]A^2=f^2
    \label{eq:duffing_plot}
\end{align}
in the rotating wave approximation (RWA), where $\delta\Omega_\text{d}=\Omega_\text{d}-(\Omega_\text{m}+\alpha)$ is the mechanical driving detuning. 
We can rewrite it as:
\begin{align}
    \frac{9}{16}\tilde{\beta}^2A^6-\frac{3}{2}\tilde{\beta}\tilde{\delta}A^4 +\left(\tilde{\delta}^2+\frac{1}{4}\right)A^2=\tilde{f}^2,
\end{align}
where we introduced damping-normalized parameters
\begin{align}
    \tilde\delta=\frac{\delta\Omega_\text{d}}{\Gamma}, \hspace{0.5cm}\tilde\beta=\frac{\beta}{\Gamma}, \hspace{0.5cm}\tilde f=\frac{f}{\Gamma}.\nonumber\\
    \label{eq:normalized_by_Gamma}
\end{align}

\begin{figure}
    \centering
    \includegraphics[width=0.5\linewidth]{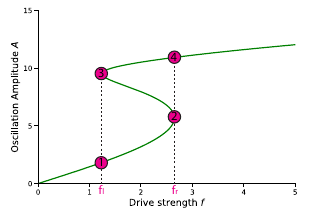}
    \caption{{\bf Characteristic points of a standard Duffing response curve.}  Oscillation amplitudes denoted $A_1$ to $A_4$. The result is computed using Eq.~\ref{eq:duffing_A} with coefficients $\Omega_\text{m}=1$, $\beta=0.01$ and $\Gamma=0.1$. 
    }
    \label{fig:SI_theory_duffing_ratio}
\end{figure}

In the hardening regime ($\beta>0$), the response becomes multistable once $\tilde{\delta} > \sqrt{3}/2$. The multistability boundaries are given by $\tilde{f}_{l}$ and $\tilde{f}_{r}$, as shown in Fig.~\ref{fig:SI_theory_duffing_ratio}:
\begin{align}
    \tilde f_{l,r}=\sqrt{\frac{18\tilde\delta+8\tilde\delta^3\mp\sqrt{(4\tilde\delta^2-3)^3}}{81\tilde\beta}}
    \label{eq:f_lr}
\end{align}
By substituting Eq.~\ref{eq:f_lr} into Eq.~\ref{eq:normalized_by_Gamma}, we obtain the corresponding four oscillation amplitudes shown in Fig.~\ref{fig:SI_theory_duffing_ratio}:
\begin{align}
    &A_{1,4}=\frac{2}{3}\sqrt{\frac{2\tilde \delta\mp\sqrt{4\tilde\delta^2-3}}{\tilde\beta}},\nonumber\\
    &A_{2,3}=\frac{2}{3}\sqrt{\frac{2\tilde \delta\mp\frac{1}{2}\sqrt{4\tilde\delta^2-3}}{\tilde\beta}}.
\end{align}

To avoid using the hysteresis region as part of the input range, the boundaries of the hysteresis window must satisfy
\begin{align}
    f_l &> (1-R_0)B, \\
    f_r &< 2(1-R_1)B,
\end{align}
where $f_l$ and $f_r$ denote the left and right boundaries of the hysteresis region, respectively. 
This leads to the constraint
\begin{align}
    \frac{f_r}{f_l} 
    < \frac{2(1-R_1)}{1+R_0}
    < 2 .
\end{align}
Substituting this result into Eq.~\ref{eq:f_lr} yields
\begin{equation}
    \tilde{\delta} < 2.47 .
    \label{eq:delta_constraint}
\end{equation}

We emphasize that satisfying Eq.~\ref{eq:delta_constraint} is a necessary condition for realizing an AND gate, but it is not sufficient; meeting this inequality alone does not guarantee correct AND gate operation, as that also requires that all inequalities Eqs.~\ref{eq:casc1}-\ref{eq:casc4} are satisfied.

\subsubsection{Cascading constraint with standard Duffing nonlinearity}
\begin{figure}
    \centering
    \includegraphics[width=\linewidth]{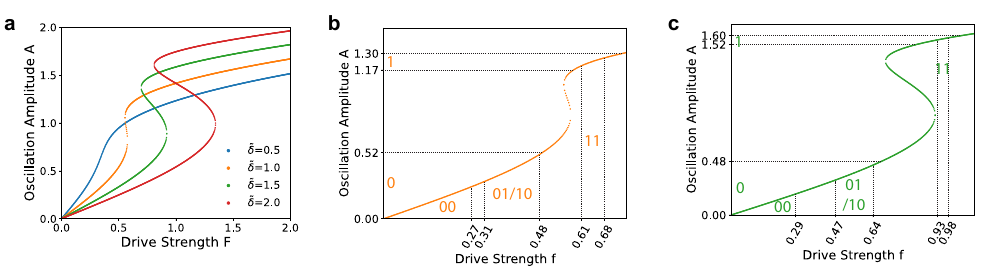}
    \caption{{\bf AND gates implemented using a standard Duffing nonlinear response that satisfies the cascading constraint.}
    {\bf a.} Duffing response curves for different detunings $\tilde\delta$. 
    The cases $\tilde\delta=1$ and $\tilde\delta=1.5$ satisfy the cascading constraint, whereas a too small detuning ($\tilde\delta=0.5$) or an large detuning  ($\tilde\delta=2$) fails to meet the requirement. 
    The Duffing response is calculated from Eq.~\ref{eq:duffing_plot} with $\Omega_m=1$, $\beta=1$, and $\Gamma=1$. 
    {\bf bc.} Comparison of AND gates constructed from a standard Duffing nonlinear response for $\tilde\delta=1$ b and $\tilde\delta=1.5$ c. 
    The hysteresis window at $\tilde\delta=1$ is smaller than that at $\tilde\delta=1.5$, allowing larger values of $R_0$ and $R_1$ for AND gate implementation, which improves noise tolerance. 
    The plots indicate the input ranges corresponding to 00, 01/10, and 11, as well as the output ranges for bits~0 and~1.}
    \label{fig:SI_cascading constraint_duffing}
\end{figure}

In this section, we investigate what nonlinear response is most suitable, focusing on the AND gate as a paradigmatic example.
Keeping all other parameters of the Duffing nonlinearity fixed, we examine how the response curve evolves with different drive detunings $\tilde\delta$, as shown in Fig.~\ref{fig:SI_cascading constraint_duffing}a. 
When the detuning $\tilde\delta$ is small, the response curve exhibits no $S$-shaped hysteresis region. 
As the detuning increases, a pronounced hysteresis window emerges and becomes progressively wider.

We numerically scan the full four-dimensional parameter space defining the logic gate, as introduced in Sec.~\ref{sec:cascading constraint}, namely $B$, $C$, $R_0$, and $R_1$. 
For $\tilde\delta=0.5$ and $\tilde\delta=2$, we find that no combination of these four parameters satisfies the cascading constraints required for implementing an AND gate. 
This conclusion also holds for other Duffing responses with different $\beta$, $\Gamma$, and $\Omega_\mathrm{m}$: if the hysteresis of the mechanical response curve is absent, too small, or so large that it affects the input ranges, the response cannot be used to realize an AND gate under the cascading constraint.

In contrast, for $\tilde\delta=1$ and $\tilde\delta=1.5$, multiple parameter combinations $(B, C, R_0, R_1)$ satisfy the cascading conditions. 
For each case, we present one representative selection in Fig.~\ref{fig:SI_cascading constraint_duffing} b and c. 
Notably, the smaller hysteresis region (yellow curve) allows larger admissible input and output bit ratios $R_0$ and $R_1$. 
This is advantageous for practical logic-gate implementation, since larger bit margins improve robustness against unavoidable experimental noise and limited precision in parameter control.

\subsubsection{Comparison between theory and experiment}
In the previous section, we discussed the idealized case described by the simplified zero-temperature Duffing model, where noise effects were neglected.
In experiments, however, unavoidable noise sources exist, as described in Sec.~\ref{sec: noise and fidelity}. 
Therefore, a more strict cascading criterion is required in practice. 
From the previous analysis, we learned that the hysteresis region should be sufficiently small while maintaining a reasonable gap between the two branches. 
Experimentally, however, we observe that the gap is larger than predicted by the simplified theory, as shown in Fig.~\ref{fig:SI_cascading constraint_duffing} and Fig.~\ref{fig:SI_jump_repeat50}b. 
We consider two possible reasons for this deviation from the idealized Duffing theory:

First, while in the ideal Duffing model the system is expected to jump from $A_2$ to $A_4$ only when the driving strength reaches the critical force $f_l$ during a forward sweep, 
in practice thermal fluctuations and the intrinsic instability of the upper branch near the bifurcation point can cause jumps to occur earlier than the deterministic threshold, effectively enlarging the observed gap.

Second, higher-order nonlinearities neglected in our simplified model, such as quadratic and quartic terms, can modify the response curve and alter the jump conditions. 
We neglect these terms in the analytical treatment because they do not constitute the essential mechanism of Duffing nonlinearity and would obscure the physical insight of the simplified model. By comparing the measured response versus drive strength in the main text Fig.~1c to the corresponding idealized Duffing response (e.g. the orange curve in Fig.~\ref{fig:SI_cascading constraint_duffing}a, with comparable hysteresis area), we observe that the slopes of amplitude vs drive strength both before and after the gap are steeper in the idealized Duffing theory than in the experiment. This means that fluctuation-induced jumps cannot fully account for the observed gap. Instead, we hypothesize that the specific higher-order character of the optomechanically-induced mechanical nonlinearity leads to a larger gap and more pronounced threshold behavior than a simple Duffing resonator. It will be interesting to study in more depth how the nonlinear response can be maximized through system and control parameters to optimize computational functions.

\subsection{Thermal noise and expected error rates}
\label{sec: noise and fidelity}
In Fig.~4c of the Main Text, we present three curves that predict the error rate of a digital gate from independent observations. 
Noise has several contributions, including thermal noise, shot noise, and slow drifts of the resonance frequency arising from low-frequency technical phase noise. To account for these effects, we introduce different curves at the beginning of this section, each corresponding to a different noise contribution included in the prediction. In this way, the individual noise sources can be distinguished and analyzed separately.
The following describes details on the theory of those prediction curves.\\
\\
The signal from the photodiode is fed into a lock-in amplifier which demodulates the signal. A lock-in amplifier is used because it selectively detects only the signal at the reference frequency. The demodulation creates two quadrature components, X (in-phase) and Y (out-of-phase), which are passed through a low-pass filter with a configurable bandwidth. The output is a complex signal $V = X + iY$ of which we record timetraces. We can reconstruct the unfiltered power spectral density from the lock-in amplifier output signal $V(t)$, since the function of the lowpass filter is known. spectral shape of the low-pass filter is known. Different filter functions can then be applied afterward to evaluate the signal fidelity for various filter bandwidths.
This results in spectra as shown in Fig.~4b for either continuously driven or undriven states. At the edges of the spectrum the flat noise floor creeps up due to a slight difference in the filter and compensation.

To establish predicted error rates from such measured fluctuations it is useful to distinguish the undriven (input 00) and driven (inputs 01, 10 and 11) cases. For each input we record a single timetrace (with identical filter bandwidths) of the output signal $V(t)$. Based on those timetraces we can calculate the expected probability distributions of the output power for different filter bandwidths. Below we describe these calculations in detail. Finally, we describe how we calculate the expected error rates, based on those distributions, and we also discuss how the error rates change when the filter bandwidth becomes smaller than the bit rate.

\subsubsection{Undriven oscillator}
\label{sec:Undriven oscillator}
\begin{figure}[]
    \centering
    \includegraphics[width=\linewidth]{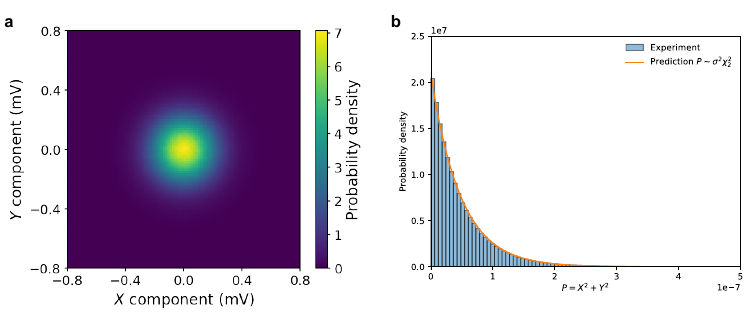}
    \caption{{\bf Distribution of the output signal $V(t)=X+iY$ when the signal-laser intensity is not modulated.} {\bf a.} Measured distribution of the $X$ and $Y$ components of the experimental output signal. {\bf b.} Experimental distribution of the signal power $P=X^2+Y^2$ and the theoretical prediction based on Eq.~\ref{eq:zero_gaussian}, where the filter bandwidth is 10 kHz.}
    \label{fig:fidelity_bit0}
\end{figure}

In the absence of a drive (no input signal), both quadratures perform a random walk in phase space due to thermal noise. For reasonable filter bandwidths, thermal noise dominates the output signal over shot noise. Both $X$ and $Y$ quadratures can be modeled as independent, zero-mean Gaussian random variables with equal variance:
\begin{align}
    X\sim \mathcal{N}(0,\sigma^2), \hspace{0.3cm} Y\sim\mathcal{N}(0,\sigma^2),
    \label{eq:zero_gaussian}
\end{align}
where $\sigma^2$ is variance of $X$ and $Y$. To confirm our experimental signals satisfy this, we record a timetrace of the undriven mode and show the corresponding joint probability distribution in Fig.~\ref{fig:fidelity_bit0}a as a function of $X$ and $Y$. The measured probability distribution is a symmetric 2D Gaussian as expected. 
Additionally, we analyze the signal power $P=X^2+Y^2$, which follows a scaled chi-square distribution for two degrees of freedom:
\begin{equation}
    \frac{P}{\sigma^2}\sim\chi_2^2.
\end{equation}
The mean and the standard deviation of $P$ have the same value:
\begin{equation}
    \mathrm{E}(P)=\mathrm{Std}(P)=2\sigma^2.
    \label{eq:sigma_bit1}
\end{equation}
Therefore, measuring $\sigma^2$ is sufficient to predict the  distribution of $P$. We compute $\hat{\mathrm{E}}(P) = 2\sigma^2$ from the recorded timetrace and predict the distribution of $P \sim \sigma^2 \chi_2^2$. We compare the predicted distribution with the empirical distribution obtained directly from the timetrace $V(t)$, as shown in Fig.~\ref{fig:fidelity_bit0}b. The two distributions are in excellent agreement. To plot the full prediction curve in Fig.~4 of the Main Text, we need the variance $\sigma^2$ for different filter bandwidths. Rather than measuring a new time trace $V(t)$ for each bandwidth, we calculate the expected variance $\sigma^2$ for different filter bandwidths based on a single timetrace. First we calculate the power spectral density (PSD) based on the timetrace. Then we divide by the spectral shape of the low-pass filter to obtain an effective unfiltered PSD. The integral of the PSD equals the mean variance of the timetrace. So by applying low-pass filters with different bandwidths to the PSD and subsequently integrating, we extract $\sigma^2$ as a function of bandwidth. Since the probability distribution of $P$ only depends on $\sigma^2$, we have now obtained the probability distributions for all filter bandwidths. It is worth noting that this prediction includes both thermal noise and shot noise. But for all applied bandwidths the thermal noise dominates over shot noise, as can be seen in Fig.~4b of the Main Text.

\subsubsection{Driven oscillator (Spectrum-based prediction)}
\label{sec:Driven oscillator (Spectrum-base prediction)}
\begin{figure}[]
    \centering
    \includegraphics[width=\linewidth]{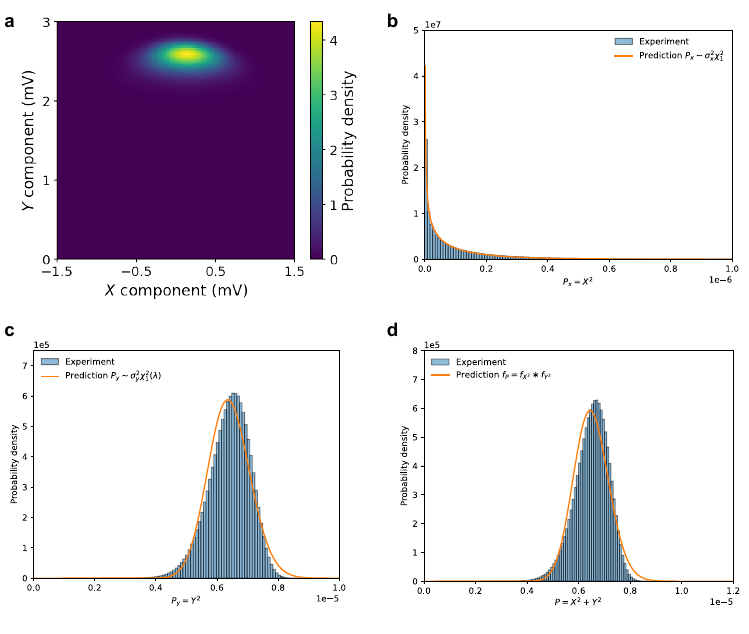}
    \caption{{\bf Distribution of the output signal $V(t)=X+iY$ when the nanobeam in driven under input 11.} {\bf a.} Distribution of the rotated $X$ and $Y$ components of the experimental output signal. {\bf b, c.} Experimental distribution of the signal power $P_x=X^2$ and $P_y=Y^2$ and the theoretical prediction based on Eq.~\ref{eq:bit1_P}, where the filter bandwidth is 10 kHz. {\bf d.} Experimental distribution of the signal power $P=X^2+Y^2$ and the theoretical prediction using the numerical convolution $f_P=f_{X^2}*f_{Y^2}$, where $f$ indicates probability density function.}
    \label{fig:fidelity_bit1}
\end{figure}

When the mode is driven by the input signal lasers (input 01, 10, and 11), the joint distribution of the two quadratures $(X,Y)$ changes with respect to the undriven case. In a linear system with noiseless drive, the drive is expected only to displace the original distribution, which would result in a Gaussian distribution with non-zero mean. We record two timetrace of the complex output signal $V(t)$ using inputs 10 and 11 and plot the joint probability distribution of the timetrace under input 11 in Fig.~\ref{fig:fidelity_bit1}a. We indeed observe that the distribution is displaced from 0, but the shape is no longer a symmetric Gaussian. Instead, the distribution is elongated along a circular arc. We interpret this shape as the combination of a nonlinearly transduced thermal noise distribution and phase fluctuations of the oscillator. A possible origin of the phase fluctuations is fluctuations of the resonance frequency \cite{fong2012frequency}. We hypothesize two mechanisms can contribute to this behavior. First, because the Duffing frequency shift depends on the oscillation amplitude, thermal-noise–induced amplitude fluctuations translate into fluctuations of the instantaneous resonance frequency, thereby producing excess phase fluctuations as a result of nonlinearity. Second, technical noise or intrinsic material defects can introduce additional slow fluctuations of eigenfrequency \cite{fong2012frequency}.
To remove correlations between the two quadratures, the shown $(X,Y)$ data in Fig.~\ref{fig:fidelity_bit1}a is rotated from the originally recorded data $(X_0, Y_0)$ by an angle $\theta$. This rotation is equivalent to using a different demodulator phase. The angle $\theta$ is chosen by minimizing the magnitude of the cross-covariance, i.e. ${\theta = \arg\min_{\theta}\, \left|\mathrm{Cov}(X,Y)\right|}$. Additionally, this places the mean of distribution at $X=0$. 

\begin{figure}
    \centering
    \includegraphics[width=0.9\linewidth]{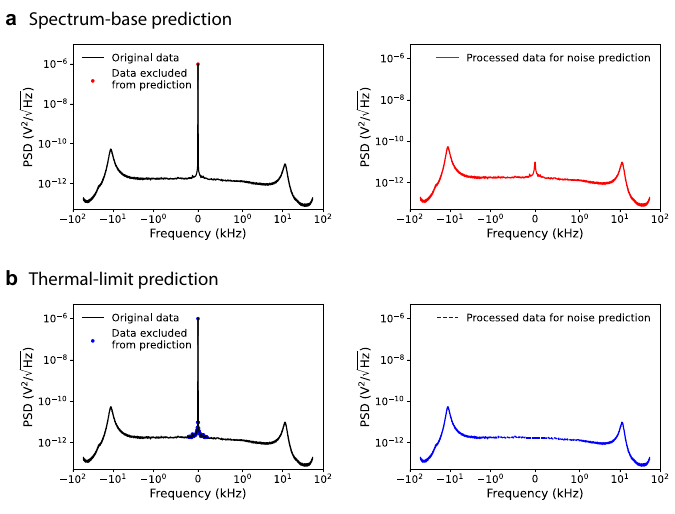}
    \caption{{\bf Noise treatment in the thermal-limit prediction (a) and the spectrum-based prediction (b).} The left panels show the original PSD data (black), together with the data points excluded from the noise analysis (red and blue markers). The right panels show the processed PSD, whose integrated area is used as the noise reference.}
    \label{fig:SI_noise_method}
\end{figure}

To simplify the error rate prediction, we approximate the joint distribution of the quadratures as a 2D Gaussian with non-zero mean and unequal variances $\sigma_X^2$ and $\sigma_Y^2$. Additionally we assume that $X$ and $Y$ are independent;
\begin{equation}
    X\sim \mathcal{N}(0, \sigma_X^2), \qquad 
    Y\sim \mathcal{N}(A,\sigma_Y^2),
\end{equation}
where $A$ is the oscillation amplitude.
Based on this approximation, $P_X=X^2$ follows a scaled chi-squared distribution with one degree of freedom, and $P_Y=Y^2$ follows a scaled non-central chi-squared distribution with one degree of freedom:
\begin{equation}
    \frac{P_X}{\sigma_X^2}\sim\chi_1^2, \qquad 
    \frac{P_Y}{\sigma_Y^2}\sim\chi_1^2(\lambda), \qquad 
    \label{eq:bit1_P}
\end{equation}
where $\lambda = A^2/\sigma_y^2$ is the noncentrality parameter. The mean of $P_X$ and $P_Y$ are:
\begin{align}
    &\mathrm{E}(P_X) = \sigma_x^2, \\
    &\mathrm{E}(P_Y) = A^2 + \sigma_y^2.
\end{align}
The prediction of $P_X$ requires a single parameter, $\sigma_X^2$, which can obtained by averaging the time trace of $X^2(t)$ or integrating over the PSD of $X$. In contrast, the prediction of $P_Y$ requires two parameters, $A^2$ and $\sigma_y^2$. Integrating over the PSD of $Y$ yields the sum of those parameters because the PSD of $Y$ contains, next to a broad thermal noise, also a sharp $\delta$-like driving peak originating from the drive. The thermal noise and sharp driving peak can also be observed in the PSD of $V$ as shown in Fig.~4b of the Main Text. We integrate over the peak to find $A^2$. To find $\sigma_Y^2$ we remove the sharp peak and integrate over the whole spectrum. Note that there are extra thermal peaks visible in the PSD at approximately $12.5~\mathrm{MHz}$ and $-25~\mathrm{MHz}$, due to non-linear mixing in the detection. In this method these extra peaks are accounted for in the prediction and contribute to the expected $\sigma_Y^2$. Based on the 3 extracted parameters we predict the probability distribution of $P_X$ and $P_Y$. We also compute the experimental distributions of $P_X$ and $P_Y$ based on the recorded timetrace (after rotation) and compare them to the expected predictions in Figs.~\ref{fig:fidelity_bit1}b and c. We again find excellent agreement between expectation and experiment for the distribution of $P_X$. However, the slight discrepancies between experimental and predicted distributions of $P_Y$ we attribute to the fact that the $(X,Y)$ distribution is only approximately Gaussian. We find the expected distribution of $P = P_X + P_Y$ by numerically convolving the expected distributions of $P_X$ and $P_Y$. We show the comparison of the expected distribution of $P$ with the experimental distribution directly extracted from the recorded timetrace in Fig.~\ref{fig:fidelity_bit1}d. Using the same method as in Sec.~\ref{sec:Undriven oscillator}, we find the expected distributions for different filter bandwidths by applying a numerical filter to the PSD. Next we obtain the parameters $\sigma_X^2$, $\sigma_Y^2$, and $A^2$ for each filter bandwidth and calculate the expected probability distribution of $P$.

\subsubsection{Driven oscillator (Thermal-limit prediction)}
\label{sec:Driven oscillator (Thermal-limit prediction)}
In the previous section, the prediction of the error rate incorporates the full measured noise spectrum under drive. 
As shown in Fig.~\ref{fig:SI_noise_method}a, significant excess noise appears in the vicinity of the narrow, delta-like peak, which should not be attributed to purely thermal fluctuations. 
This contribution becomes particularly relevant when the bandpass filter bandwidth is small. 
By excluding the noise around the narrow peak in the present analysis, we can isolate the role of thermal noise in limiting the computation fidelity. 

The procedure follows Sec.~\ref{sec:Driven oscillator (Spectrum-base prediction)}. 
The only difference is that we now extract $\sigma_x$ and $\sigma_y$ after removing all data points around the central peak from the spectrum, as indicated by the blue markers in Fig.~\ref{fig:SI_noise_method}b, left panel. 
The excluded contribution is then compensated using the plateau level of the broadband background to maintain consistent normalization, as shown in Fig.~\ref{fig:SI_noise_method}b, right panel.

As shown in Fig.~4b of the Main Text, two Lorentzian peaks are visible in the driven spectrum. 
The left peak corresponds to the original thermal resonance, shifted from the undriven frequency due to Duffing softening. 
The right peak arises from nonlinear mixing between the thermal Lorentz resonance and the narrow peak associated with the driven response.

\begin{figure}[]
    \centering
    \includegraphics[width=0.5\linewidth]{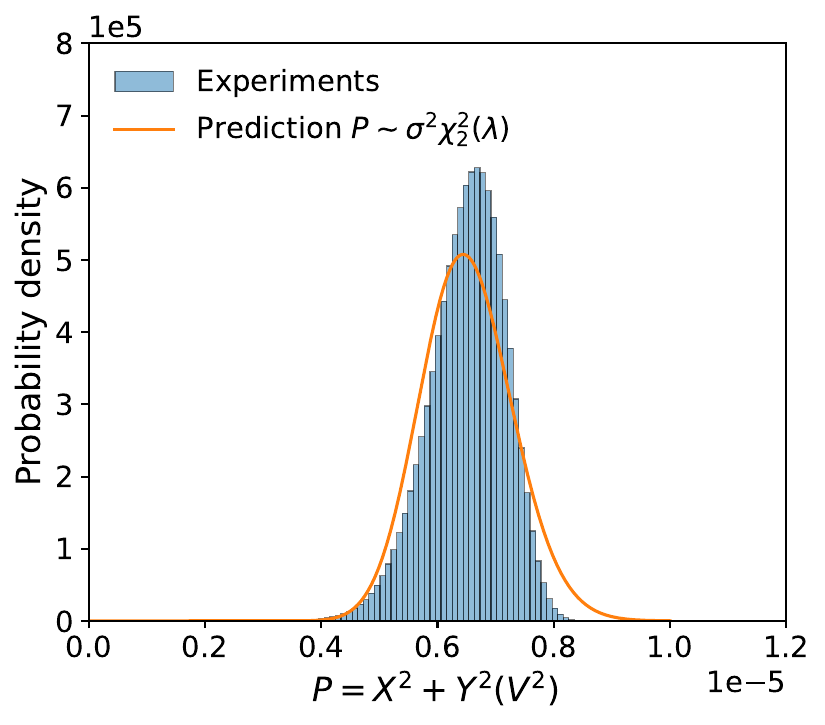}
    \caption{{\bf Predicted signal distribution of the driven nanobeam assuming purely thermal noise.}}
    \label{fig:fidelity_bit1_thermal}
\end{figure}

\subsubsection{Driven oscillator (Thermal-limit prediction based on undriven noise)}
\label{sec:Driven oscillator (Thermal-limit prediction based on undriven noise)}
Here, for reference we adopt an alternative approach to predict the computation error rate induced by thermal noise. 
Specifically, we estimate the error rate of the driven mode using noise statistics extracted from a time trace of the undriven mode.

Our model for predicting $P_X$ and $P_Y$ remains the same and is based on Eq.~\ref{eq:bit1_P}. We assume that the only effect of the driving is the addition of oscillation amplitude $A$ and that the variance of the thermal noise $\sigma^2$ equals the variance of the undriven case. Thus we obtain $\sigma^2$ from the undriven data using Eq.~\ref{eq:sigma_bit1}. We then set $\sigma_x=\sigma_y=\sigma$ in Eq.~\ref{eq:bit1_P}. It follows that
\begin{equation}
    \frac{P_x}{\sigma^2}\sim \chi_1^2, \qquad 
    \frac{P_y}{\sigma^2}\sim \chi_1^2(\lambda).
\end{equation}
Therefore, $P=P_x+P_y$ follows
\begin{align}
    \frac{P}{\sigma^2} &\sim \chi_2^2(\lambda),\\
    \mathrm{E}(P) &= A^2+2\sigma^2.
\end{align}
To predict the non-centrality parameter $\lambda=A^2/\sigma^2$, we obtain $A^2$ by combining the variance of the undriven case $\sigma^2$ with $\mathrm{E}(P)$ of the driven timetrace, either be averaging the timetrace of integrating over the PSD of the output signal $V$. We plot the expected distribution of $P$ in Fig.~\ref{fig:fidelity_bit1_thermal} together with the experimental distribution based on the driven timetrace. The calculation of the expected distributions for different filter bandwidths is again based on applying different low-pass filter to the (undriven) spectra.

It is worth noting that, as can be seen from the thermal noise in Fig.~4b of the Main Text, the effective eigenfrequency of the driven case is shifted compared to the undrive case, due to the Duffing nonlinearity. We therefore shift the undriven PSD along the frequency axis before applying the low-ass filter, such that the undriven resonance is at the same frequency as the driven resonance.

\subsubsection{Error rate prediction over different filter bandwidths}
There are four possible input states, 00, 01, 10, and 11. In our experiment, 01 and 10 are equivalent because the two signal lasers are calibrated to produce identical inputs. For each input state, we use the predicted output distribution together with the predefined thresholds for output bits 0 and 1 to compute the probability that the output falls outside the correct decision range, which defines the error rate for that input. It is worth noting that, for an OR gate, although both inputs 01 (10) and 11 ideally produce output bit 1, they correspond to different drive conditions and therefore lead to different output distributions. We thus evaluate their error contributions separately. For a random input sequence in which the four input states occur with equal probability, the overall error rate is then given by the average over the four cases.

\begin{figure}[]
    \centering
    \includegraphics[width=0.5\linewidth]{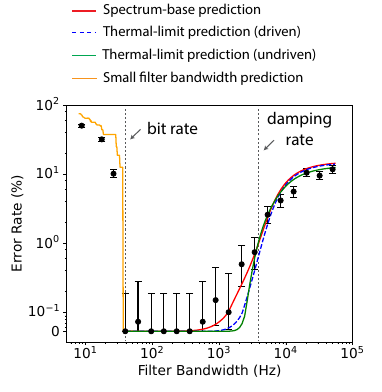}
    \caption{{\bf Predictions of the computation error rate based on different models.} Black scatter points and error bars show the experimental data and its confidence interval.}
\label{fig:SI_fidelity_greencurve}
\end{figure}

For the ``Spectrum-based prediction (driven)'' curve (red), we again treat input 00 using the method in Sec.~\ref{sec:Undriven oscillator}. For the driven input states 01(10) and 11, and 11, we use the spectrum-based prediction described in Sec.~\ref{sec:Driven oscillator (Spectrum-base prediction)}, which incorporates all the measured noise spectrum under drive.

For the ``Thermal-limit prediction'' curve (blue dashed line) in Fig.~4c of the Main Text, we compute the error rate for input 00 using the method in Sec.~\ref{sec:Undriven oscillator}. 
For the driven input states 01(10) and 11, we use the thermal-noise limit inferred from the undriven measurement, following the procedure described in Sec.~\ref{sec:Driven oscillator (Thermal-limit prediction)}.

The ``Thermal-limit prediction (undriven)'' curve is not shown in Fig.~4c of the Main Text. 
For comparison, we plot it in Fig.~\ref{fig:SI_fidelity_greencurve} as the green curve.
We compute the error rate for input 00 using the method in Sec.~\ref{sec:Undriven oscillator}. 
For the driven input states 01(10) and 11, we use the thermal-noise limit inferred from the undriven measurement, following the procedure described in Sec.~\ref{sec:Driven oscillator (Thermal-limit prediction based on undriven noise)}.

For the ``Small filter bandwidth prediction'' curve (yellow), we use a different approach. 
When the bit rate exceeds the filter bandwidth, the lock-in output effectively averages over the previous few bit periods. 
We model this effect numerically by generating a random input sequence with $10^5$ data points, predicting the corresponding ideal output sequence, and then averaging over the previous few bit periods according to the filter bandwidth. Since the filter bandwidth is very small in this regime, the contribution from thermal noise to the error rate becomes negligible compared to the effect of averaging over multiple bit periods. We therefore neglect noise contributions and use the averaged measured output levels as deterministic values in the prediction. It is worth noting that, although the programmed bit rate in the experiment is $100~\mathrm{Hz}$, the effective bit rate is lower due to the response time of the control electronics and software overhead. Based on the experimental data, we take the effective bit rate to be $40~\mathrm{Hz}$. Below $40~\mathrm{Hz}$ the error rate increases sharply, indicating that the lock-in readout averages over multiple preceding bits. This is mainly due to the control method used for communication between the scripts and the lock-in amplifier. The performance can be improved by implementing a more efficient control approach.

\section{Supplementary Figures}

\begin{figure*}[!htbp]
    \centering
    \includegraphics[width=0.5\linewidth]{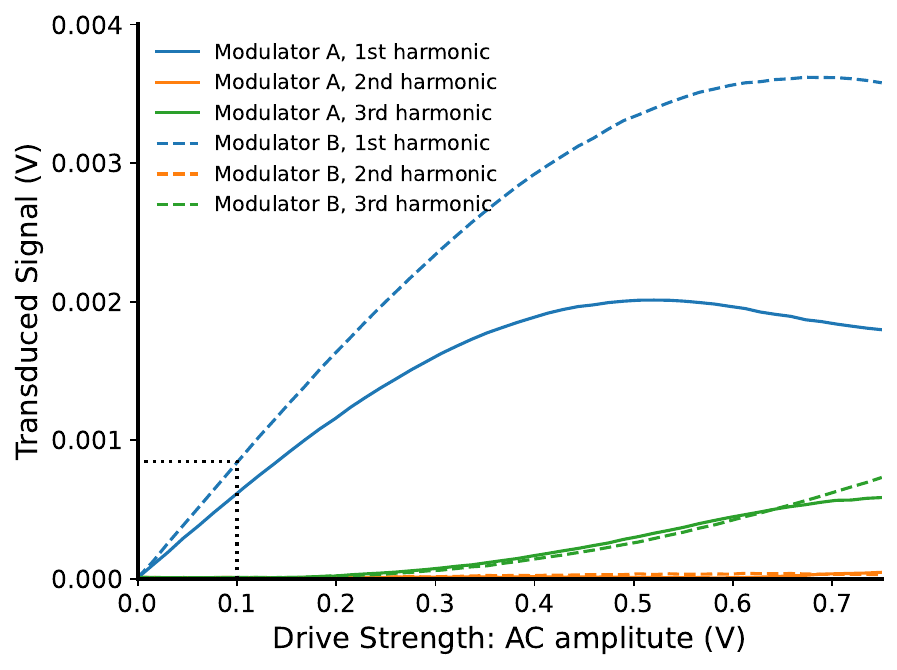}
    \caption{{\bf Nonlinearity of intensity modulators.} The transduced modulation strength is not a linear function of the input modulation amplitude. In our experiments we operate within 0–0.1 V (the dashed-box region), where the response is approximately linear. The optimized DC biases are 2.0 V for Modulator A and 3.3 V for Modulator B.}
    \label{fig:SI_modulator_linear}
\end{figure*}

\begin{figure}[!htbp]
    \centering
    \includegraphics[width=0.75\linewidth]{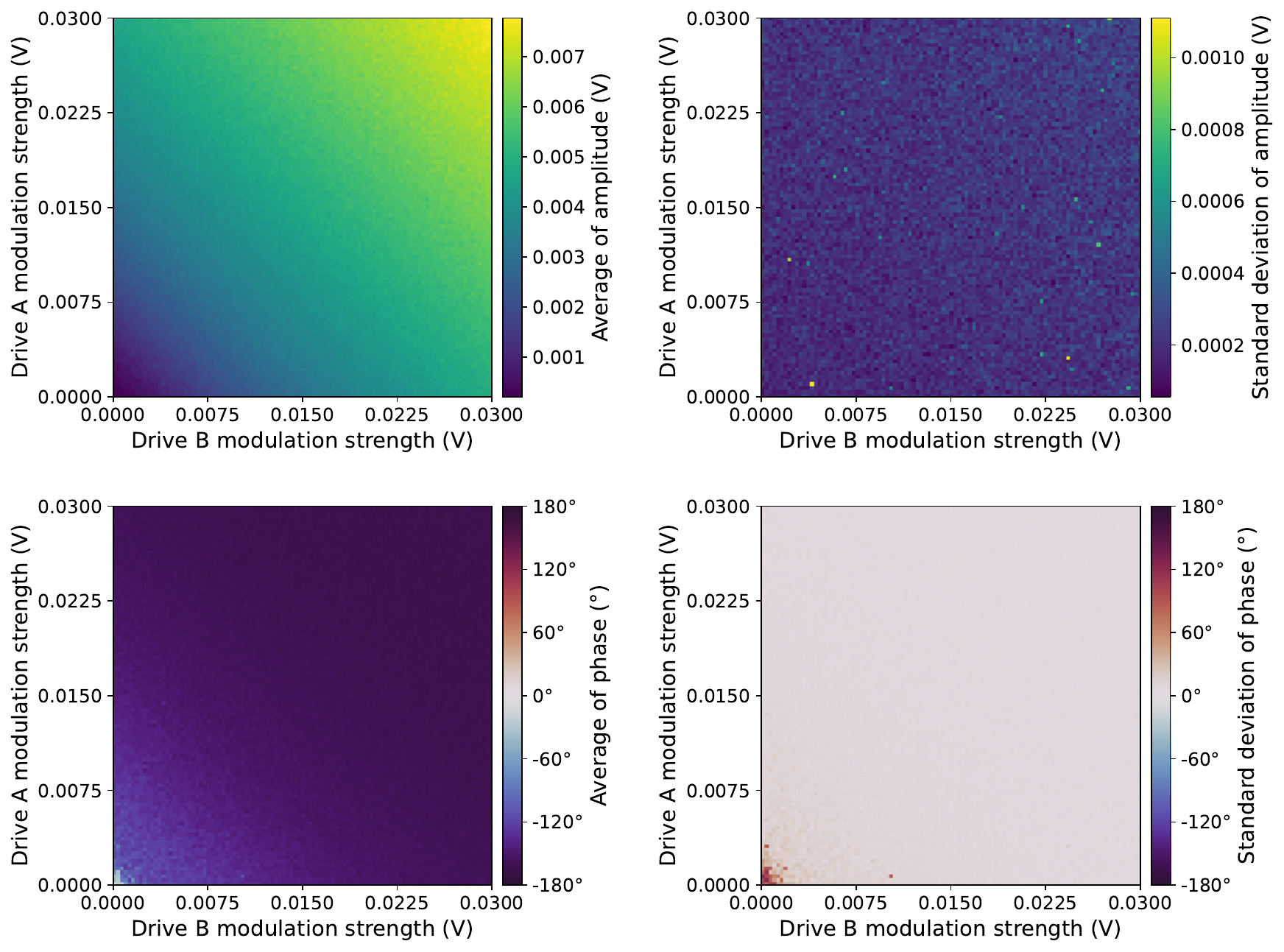}
    \caption{{\bf Average and standard deviation of two-driving sweep results over 10 repeats, with the function laser off.}}
    \label{fig:SI_linear_logic_repeat}
\end{figure}
\begin{figure}[h!]
    \centering
    \includegraphics[width=0.75\linewidth]{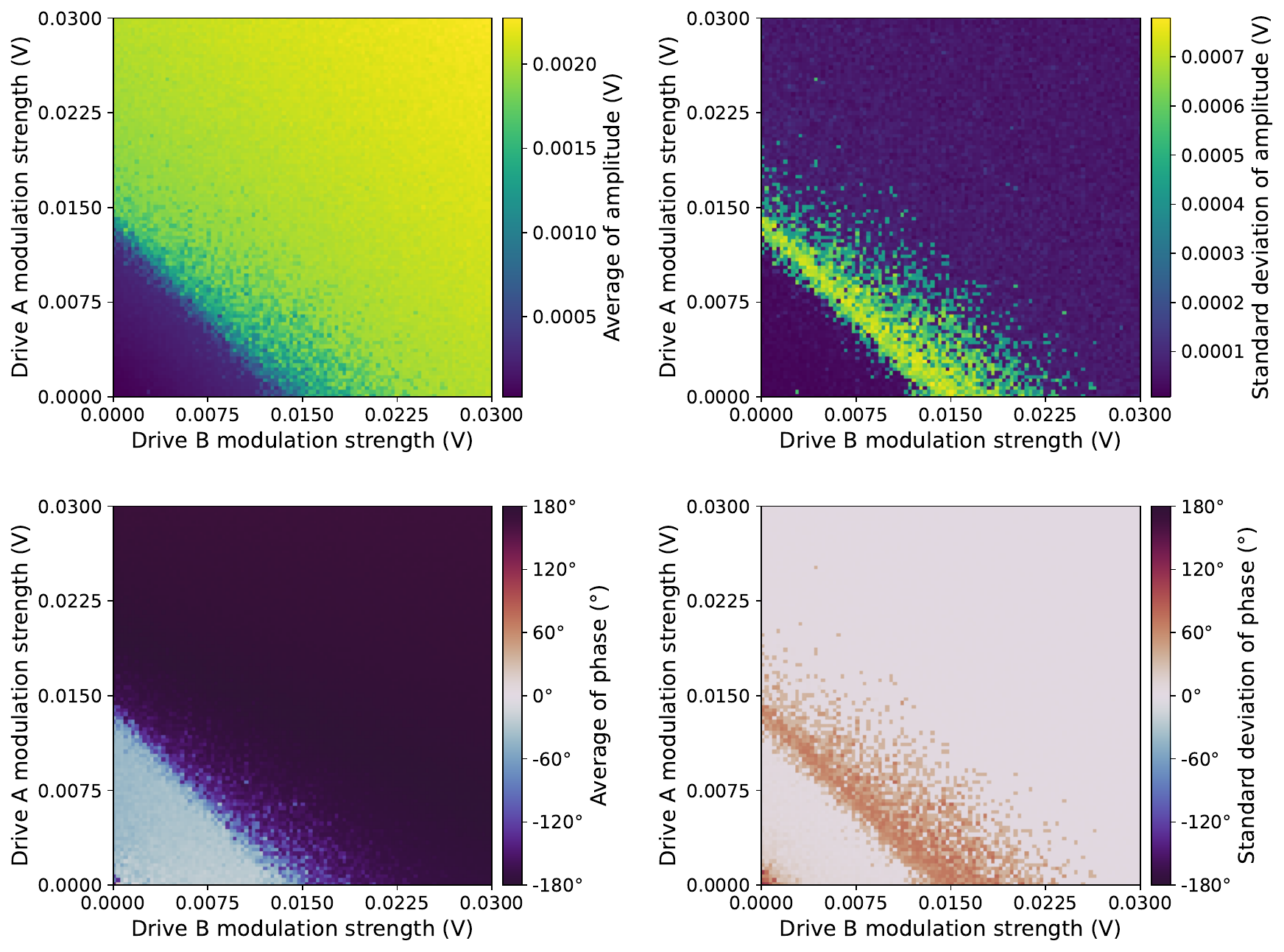}
    \caption{{\bf Average and standard deviation of two-driving sweep results over 10 repeats, with the function laser on.}}
    \label{fig:SI_nonlinear_logic_repeat}
\end{figure}

\begin{figure*}[!htbp]
    \centering
    \includegraphics[width=\linewidth]{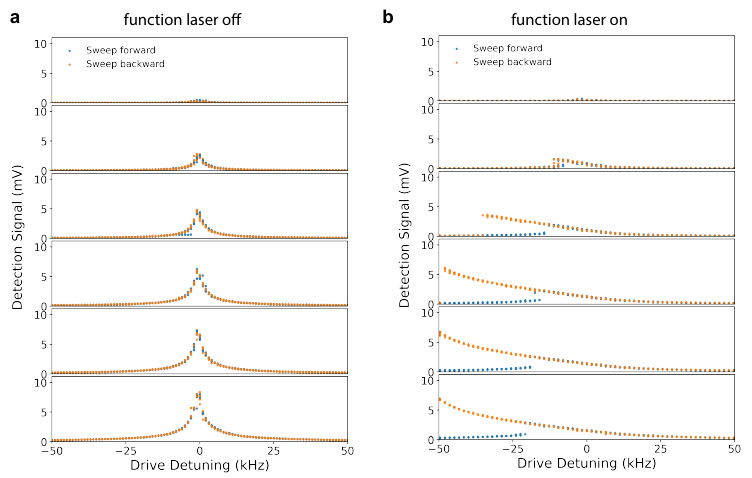}
    \caption{{\bf Frequency response of the optomechanical system as the drive strength increases.} {\bf a.} With the function laser off, the response follows the square-root Lorentzian line shape. {\bf b.} With the function laser on, the response exhibits a shark-fin profile indicative of Duffing nonlinearity. Ten repeated experimental measurements are overlaid for each set of control parameters. The drive modulation strength is increased from the top to the bottom panels, corresponding to $\{0, 0.012, 0.024, 0.036, 0.048, 0.06\}$ V.
    }
    \label{fig:SI_duffing_increase_driveamp}
\end{figure*}
\newpage
\begin{figure*}[!htbp]
    \centering
    \includegraphics[width=\linewidth]{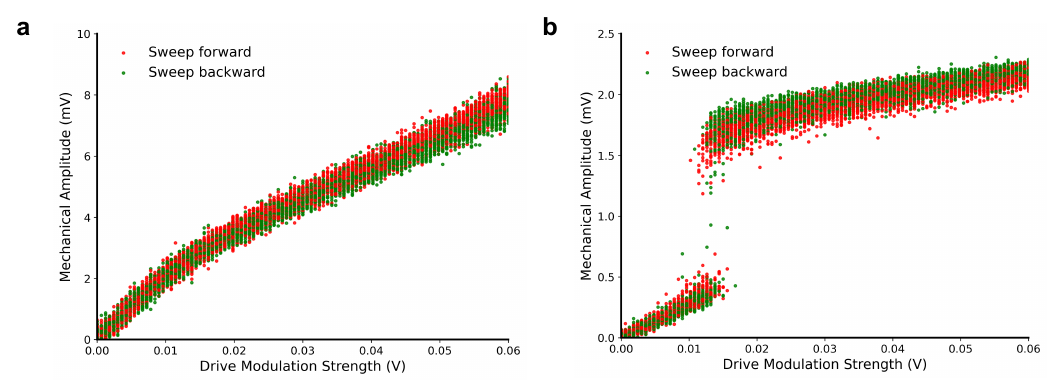}
    \caption{{\bf Mechanical amplitude of nanobeam oscillations.} {\bf a.} With the functio laser off, the mechanical amplitude increases approximately linearly with the drive modulation strength. {\bf b.} With the function laser on, the amplitude shows strong nonlinearity versus drive modulation strength. Fifty repeated measurements are overlaid to demonstrate robust and reproducible results.
    }
    \label{fig:SI_jump_repeat50}
\end{figure*}

\begin{figure*}[!htbp]
    \centering
    \includegraphics[width=\linewidth]{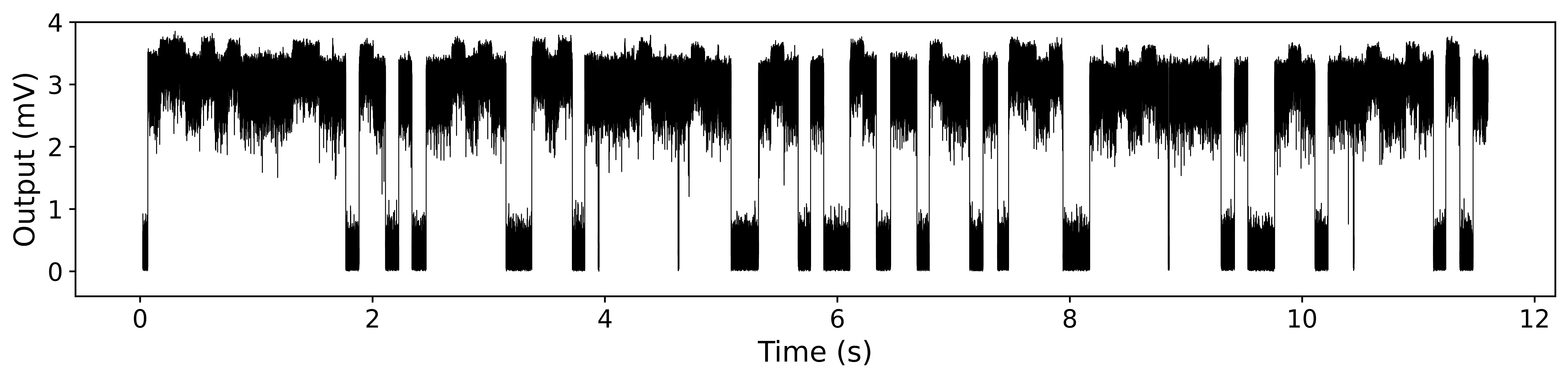}
    \caption{ {\bf Full time trace of an OR gate output under a continuous input sequence.} Since inputs 01 (10) and 11 both produce output bit 1, the output exhibits two mean levels. Data are acquired with the lock-in amplifier filter bandwidth set to 10 kHz.  }
    \label{fig:SI_longtimetrace}
\end{figure*}


\begin{thebibliography}{10}
\expandafter\ifx\csname url\endcsname\relax
  \def\url#1{\texttt{#1}}\fi
\expandafter\ifx\csname urlprefix\endcsname\relax\def\urlprefix{URL }\fi
\providecommand{\bibinfo}[2]{#2}
\providecommand{\eprint}[2][]{\url{#2}}

\bibitem{roukes2004mechanical}
\bibinfo{author}{Roukes, M.~L.}
\newblock \bibinfo{title}{Mechanical compution, redux?}
\newblock In \emph{\bibinfo{booktitle}{IEEE International Electron Devices Meeting: IEDM Technical Digest}}, \bibinfo{pages}{539--542} (\bibinfo{organization}{IEEE}, \bibinfo{year}{2004}).

\bibitem{vandoorne2014experimental}
\bibinfo{author}{Vandoorne, K.} \emph{et~al.}
\newblock \bibinfo{title}{Experimental demonstration of reservoir computing on a silicon photonics chip}.
\newblock \emph{\bibinfo{journal}{Nat. Commun.}} \textbf{\bibinfo{volume}{5}}, \bibinfo{pages}{3541} (\bibinfo{year}{2014}).

\bibitem{yasuda2021mechanical}
\bibinfo{author}{Yasuda, H.} \emph{et~al.}
\newblock \bibinfo{title}{Mechanical computing}.
\newblock \emph{\bibinfo{journal}{Nature}} \textbf{\bibinfo{volume}{598}}, \bibinfo{pages}{39--48} (\bibinfo{year}{2021}).

\bibitem{kaspar2021rise}
\bibinfo{author}{Kaspar, C.}, \bibinfo{author}{Ravoo, B.~J.}, \bibinfo{author}{van~der Wiel, W.~G.}, \bibinfo{author}{Wegner, S.~V.} \& \bibinfo{author}{Pernice, W. H.~P.}
\newblock \bibinfo{title}{The rise of intelligent matter}.
\newblock \emph{\bibinfo{journal}{Nature}} \textbf{\bibinfo{volume}{594}}, \bibinfo{pages}{345--355} (\bibinfo{year}{2021}).

\bibitem{wright2022deep}
\bibinfo{author}{Wright, L.~G.} \emph{et~al.}
\newblock \bibinfo{title}{Deep physical neural networks trained with backpropagation}.
\newblock \emph{\bibinfo{journal}{Nature}} \textbf{\bibinfo{volume}{601}}, \bibinfo{pages}{549--555} (\bibinfo{year}{2022}).

\bibitem{mcmahon2023physics}
\bibinfo{author}{McMahon, P.~L.}
\newblock \bibinfo{title}{The physics of optical computing}.
\newblock \emph{\bibinfo{journal}{Nat. Rev. Phys.}} \textbf{\bibinfo{volume}{5}}, \bibinfo{pages}{717--734} (\bibinfo{year}{2023}).

\bibitem{dubvcek2024sensor}
\bibinfo{author}{Dub{\v{c}}ek, T.} \emph{et~al.}
\newblock \bibinfo{title}{In-sensor passive speech classification with phononic metamaterials}.
\newblock \emph{\bibinfo{journal}{Adv. Funct. Mater.}} \textbf{\bibinfo{volume}{34}}, \bibinfo{pages}{2311877} (\bibinfo{year}{2024}).

\bibitem{romero2024acoustically}
\bibinfo{author}{Romero, E.} \emph{et~al.}
\newblock \bibinfo{title}{Acoustically driven single-frequency mechanical logic}.
\newblock \emph{\bibinfo{journal}{Phys. Rev. Appl.}} \textbf{\bibinfo{volume}{21}}, \bibinfo{pages}{054029} (\bibinfo{year}{2024}).

\bibitem{zhou2025harnessing}
\bibinfo{author}{Zhou, J.} \emph{et~al.}
\newblock \bibinfo{title}{Harnessing spatiotemporal transformation in magnetic domains for nonvolatile physical reservoir computing}.
\newblock \emph{\bibinfo{journal}{Sci. Adv.}} \textbf{\bibinfo{volume}{11}}, \bibinfo{pages}{eadr5262} (\bibinfo{year}{2025}).

\bibitem{Camsari2019p-bits}
\bibinfo{author}{Camsari, K.~Y.}, \bibinfo{author}{Sutton, B.~M.} \& \bibinfo{author}{Datta, S.}
\newblock \bibinfo{title}{p-bits for probabilistic spin logic}.
\newblock \emph{\bibinfo{journal}{Appl. Phys. Rev.}} \textbf{\bibinfo{volume}{6}}, \bibinfo{pages}{011305} (\bibinfo{year}{2019}).

\bibitem{lopez2016sub}
\bibinfo{author}{Lopez-Suarez, M.}, \bibinfo{author}{Neri, I.} \& \bibinfo{author}{Gammaitoni, L.}
\newblock \bibinfo{title}{Sub-k bt micro-electromechanical irreversible logic gate}.
\newblock \emph{\bibinfo{journal}{Nat. Commun.}} \textbf{\bibinfo{volume}{7}}, \bibinfo{pages}{12068} (\bibinfo{year}{2016}).

\bibitem{landauer1961irreversibility}
\bibinfo{author}{Landauer, R.}
\newblock \bibinfo{title}{Irreversibility and heat generation in the computing process}.
\newblock \emph{\bibinfo{journal}{IBM J. Res. Dev.}} \textbf{\bibinfo{volume}{5}}, \bibinfo{pages}{183--191} (\bibinfo{year}{1961}).

\bibitem{berut2012experimental}
\bibinfo{author}{B{\'e}rut, A.} \emph{et~al.}
\newblock \bibinfo{title}{Experimental verification of {Landauer}’s principle linking information and thermodynamics}.
\newblock \emph{\bibinfo{journal}{Nature}} \textbf{\bibinfo{volume}{483}}, \bibinfo{pages}{187--189} (\bibinfo{year}{2012}).

\bibitem{badzey2004controllable}
\bibinfo{author}{Badzey, R.~L.}, \bibinfo{author}{Zolfagharkhani, G.}, \bibinfo{author}{Gaidarzhy, A.} \& \bibinfo{author}{Mohanty, P.}
\newblock \bibinfo{title}{A controllable nanomechanical memory element}.
\newblock \emph{\bibinfo{journal}{Appl. Phys. Lett.}} \textbf{\bibinfo{volume}{85}}, \bibinfo{pages}{3587--3589} (\bibinfo{year}{2004}).

\bibitem{prakash2007microfluidic}
\bibinfo{author}{Prakash, M.} \& \bibinfo{author}{Gershenfeld, N.}
\newblock \bibinfo{title}{Microfluidic bubble logic}.
\newblock \emph{\bibinfo{journal}{Science}} \textbf{\bibinfo{volume}{315}}, \bibinfo{pages}{832--835} (\bibinfo{year}{2007}).

\bibitem{guerra2010noise}
\bibinfo{author}{Guerra, D.~N.} \emph{et~al.}
\newblock \bibinfo{title}{A noise-assisted reprogrammable nanomechanical logic gate}.
\newblock \emph{\bibinfo{journal}{Nano Lett.}} \textbf{\bibinfo{volume}{10}}, \bibinfo{pages}{1168--1171} (\bibinfo{year}{2010}).

\bibitem{mahboob2011interconnect}
\bibinfo{author}{Mahboob, I.}, \bibinfo{author}{Flurin, E.}, \bibinfo{author}{Nishiguchi, K.}, \bibinfo{author}{Fujiwara, A.} \& \bibinfo{author}{Yamaguchi, H.}
\newblock \bibinfo{title}{Interconnect-free parallel logic circuits in a single mechanical resonator}.
\newblock \emph{\bibinfo{journal}{Nat. Commun.}} \textbf{\bibinfo{volume}{2}}, \bibinfo{pages}{198} (\bibinfo{year}{2011}).

\bibitem{moon2012genetic}
\bibinfo{author}{Moon, T.~S.}, \bibinfo{author}{Lou, C.}, \bibinfo{author}{Tamsir, A.}, \bibinfo{author}{Stanton, B.~C.} \& \bibinfo{author}{Voigt, C.~A.}
\newblock \bibinfo{title}{Genetic programs constructed from layered logic gates in single cells}.
\newblock \emph{\bibinfo{journal}{Nature}} \textbf{\bibinfo{volume}{491}}, \bibinfo{pages}{249--253} (\bibinfo{year}{2012}).

\bibitem{fredkin2014}
\bibinfo{author}{Wenzler, J.-S.}, \bibinfo{author}{Dunn, T.}, \bibinfo{author}{Toffoli, T.} \& \bibinfo{author}{Mohanty, P.}
\newblock \bibinfo{title}{A nanomechanical {Fredkin} gate}.
\newblock \emph{\bibinfo{journal}{Nano Lett.}} \textbf{\bibinfo{volume}{14}}, \bibinfo{pages}{89--93} (\bibinfo{year}{2014}).

\bibitem{yao2014logic}
\bibinfo{author}{Yao, A.} \& \bibinfo{author}{Hikihara, T.}
\newblock \bibinfo{title}{Logic-memory device of a mechanical resonator}.
\newblock \emph{\bibinfo{journal}{Appl. Phys. Lett.}} \textbf{\bibinfo{volume}{105}} (\bibinfo{year}{2014}).

\bibitem{mahboob2014multimode}
\bibinfo{author}{Mahboob, I.}, \bibinfo{author}{Mounaix, M.}, \bibinfo{author}{Nishiguchi, K.}, \bibinfo{author}{Fujiwara, A.} \& \bibinfo{author}{Yamaguchi, H.}
\newblock \bibinfo{title}{A multimode electromechanical parametric resonator array}.
\newblock \emph{\bibinfo{journal}{Scientific reports}} \textbf{\bibinfo{volume}{4}}, \bibinfo{pages}{4448} (\bibinfo{year}{2014}).

\bibitem{raney2016stable}
\bibinfo{author}{Raney, J.~R.} \emph{et~al.}
\newblock \bibinfo{title}{Stable propagation of mechanical signals in soft media using stored elastic energy}.
\newblock \emph{\bibinfo{journal}{Proc. Natl. Acad. Sci.}} \textbf{\bibinfo{volume}{113}}, \bibinfo{pages}{9722--9727} (\bibinfo{year}{2016}).

\bibitem{dion2018reservoir}
\bibinfo{author}{Dion, G.}, \bibinfo{author}{Mejaouri, S.} \& \bibinfo{author}{Sylvestre, J.}
\newblock \bibinfo{title}{Reservoir computing with a single delay-coupled non-linear mechanical oscillator}.
\newblock \emph{\bibinfo{journal}{J. Appl. Phys.}} \textbf{\bibinfo{volume}{124}} (\bibinfo{year}{2018}).

\bibitem{ilyas2019cascadable}
\bibinfo{author}{Ilyas, S.}, \bibinfo{author}{Ahmed, S.}, \bibinfo{author}{Hafiz, M.~A.}, \bibinfo{author}{Fariborzi, H.} \& \bibinfo{author}{Younis, M.~I.}
\newblock \bibinfo{title}{Cascadable microelectromechanical resonator logic gate}.
\newblock \emph{\bibinfo{journal}{J. Micromech. Microeng.}} \textbf{\bibinfo{volume}{29}}, \bibinfo{pages}{015007} (\bibinfo{year}{2019}).

\bibitem{song2019additively}
\bibinfo{author}{Song, Y.} \emph{et~al.}
\newblock \bibinfo{title}{Additively manufacturable micro-mechanical logic gates}.
\newblock \emph{\bibinfo{journal}{Nat. Commun.}} \textbf{\bibinfo{volume}{10}}, \bibinfo{pages}{882} (\bibinfo{year}{2019}).

\bibitem{el2021digital}
\bibinfo{author}{El~Helou, C.}, \bibinfo{author}{Buskohl, P.~R.}, \bibinfo{author}{Tabor, C.~E.} \& \bibinfo{author}{Harne, R.~L.}
\newblock \bibinfo{title}{Digital logic gates in soft, conductive mechanical metamaterials}.
\newblock \emph{\bibinfo{journal}{Nat. Commun.}} \textbf{\bibinfo{volume}{12}}, \bibinfo{pages}{1633} (\bibinfo{year}{2021}).

\bibitem{mei2021mechanical}
\bibinfo{author}{Mei, T.}, \bibinfo{author}{Meng, Z.}, \bibinfo{author}{Zhao, K.} \& \bibinfo{author}{Chen, C.~Q.}
\newblock \bibinfo{title}{A mechanical metamaterial with reprogrammable logical functions}.
\newblock \emph{\bibinfo{journal}{Nat. Commun.}} \textbf{\bibinfo{volume}{12}}, \bibinfo{pages}{7234} (\bibinfo{year}{2021}).

\bibitem{lv2023dna}
\bibinfo{author}{Lv, H.} \emph{et~al.}
\newblock \bibinfo{title}{{DNA}-based programmable gate arrays for general-purpose {DNA} computing}.
\newblock \emph{\bibinfo{journal}{Nature}} \textbf{\bibinfo{volume}{622}}, \bibinfo{pages}{292--300} (\bibinfo{year}{2023}).

\bibitem{he2024programmable}
\bibinfo{author}{He, Q.}, \bibinfo{author}{Ferracin, S.} \& \bibinfo{author}{Raney, J.~R.}
\newblock \bibinfo{title}{Programmable responsive metamaterials for mechanical computing and robotics}.
\newblock \emph{\bibinfo{journal}{Nature Comput. Sci.}} \textbf{\bibinfo{volume}{4}}, \bibinfo{pages}{567--573} (\bibinfo{year}{2024}).

\bibitem{wang2024harnessing}
\bibinfo{author}{Wang, X.} \& \bibinfo{author}{Cichos, F.}
\newblock \bibinfo{title}{Harnessing synthetic active particles for physical reservoir computing}.
\newblock \emph{\bibinfo{journal}{Nat. Commun.}} \textbf{\bibinfo{volume}{15}}, \bibinfo{pages}{774} (\bibinfo{year}{2024}).

\bibitem{byun2024integrated}
\bibinfo{author}{Byun, J.}, \bibinfo{author}{Pal, A.}, \bibinfo{author}{Ko, J.} \& \bibinfo{author}{Sitti, M.}
\newblock \bibinfo{title}{Integrated mechanical computing for autonomous soft machines}.
\newblock \emph{\bibinfo{journal}{Nat. Commun.}} \textbf{\bibinfo{volume}{15}}, \bibinfo{pages}{2933} (\bibinfo{year}{2024}).

\bibitem{lee2024task}
\bibinfo{author}{Lee, O.} \emph{et~al.}
\newblock \bibinfo{title}{Task-adaptive physical reservoir computing}.
\newblock \emph{\bibinfo{journal}{Nat. Mater.}} \textbf{\bibinfo{volume}{23}}, \bibinfo{pages}{79--87} (\bibinfo{year}{2024}).

\bibitem{song2025heat}
\bibinfo{author}{Song, T.} \& \bibinfo{author}{Qian, L.}
\newblock \bibinfo{title}{Heat-rechargeable computation in dna logic circuits and neural networks}.
\newblock \emph{\bibinfo{journal}{Nature}} \textbf{\bibinfo{volume}{646}}, \bibinfo{pages}{315--322} (\bibinfo{year}{2025}).

\bibitem{watkins2025arbitrary}
\bibinfo{author}{Watkins, A.~A.}, \bibinfo{author}{Bordiga, G.}, \bibinfo{author}{Mu, M.}, \bibinfo{author}{Tournat, V.} \& \bibinfo{author}{Bertoldi, K.}
\newblock \bibinfo{title}{Arbitrary mechanical memory encoding via nonlinear waves in bistable metamaterials}.
\newblock \emph{\bibinfo{journal}{arXiv preprint arXiv:2508.20321}}  (\bibinfo{year}{2025}).

\bibitem{schmid2016fundamentals}
\bibinfo{author}{Schmid, S.}, \bibinfo{author}{Villanueva, L.~G.} \& \bibinfo{author}{Roukes, M.~L.}
\newblock \emph{\bibinfo{title}{Fundamentals of nanomechanical resonators}}, vol.~\bibinfo{volume}{49} (\bibinfo{publisher}{Springer}, \bibinfo{year}{2016}).

\bibitem{coulombe2017computing}
\bibinfo{author}{Coulombe, J.~C.}, \bibinfo{author}{York, M.~C.} \& \bibinfo{author}{Sylvestre, J.}
\newblock \bibinfo{title}{Computing with networks of nonlinear mechanical oscillators}.
\newblock \emph{\bibinfo{journal}{PloS One}} \textbf{\bibinfo{volume}{12}}, \bibinfo{pages}{e0178663} (\bibinfo{year}{2017}).

\bibitem{tadokoro2021highly}
\bibinfo{author}{Tadokoro, Y.} \& \bibinfo{author}{Tanaka, H.}
\newblock \bibinfo{title}{Highly sensitive implementation of logic gates with a nonlinear nanomechanical resonator}.
\newblock \emph{\bibinfo{journal}{Phys. Rev. Applied}} \textbf{\bibinfo{volume}{15}}, \bibinfo{pages}{024058} (\bibinfo{year}{2021}).

\bibitem{eichler2012strong}
\bibinfo{author}{Eichler, A.}, \bibinfo{author}{del {\'A}lamo~Ruiz, M.}, \bibinfo{author}{Plaza, J.} \& \bibinfo{author}{Bachtold, A.}
\newblock \bibinfo{title}{Strong coupling between mechanical modes in a nanotube resonator}.
\newblock \emph{\bibinfo{journal}{Phys. Rev. Lett.}} \textbf{\bibinfo{volume}{109}}, \bibinfo{pages}{025503} (\bibinfo{year}{2012}).

\bibitem{rieser2022tunable}
\bibinfo{author}{Rieser, J.} \emph{et~al.}
\newblock \bibinfo{title}{Tunable light-induced dipole-dipole interaction between optically levitated nanoparticles}.
\newblock \emph{\bibinfo{journal}{Science}} \textbf{\bibinfo{volume}{377}}, \bibinfo{pages}{987--990} (\bibinfo{year}{2022}).

\bibitem{jin2024engineering}
\bibinfo{author}{Jin, X.} \emph{et~al.}
\newblock \bibinfo{title}{Engineering error correcting dynamics in nanomechanical systems}.
\newblock \emph{\bibinfo{journal}{Sci. Rep.}} \textbf{\bibinfo{volume}{14}}, \bibinfo{pages}{20431} (\bibinfo{year}{2024}).

\bibitem{venkatesh2017implementation}
\bibinfo{author}{Venkatesh, P.}, \bibinfo{author}{Venkatesan, A.} \& \bibinfo{author}{Lakshmanan, M.}
\newblock \bibinfo{title}{Implementation of dynamic dual input multiple output logic gate via resonance in globally coupled {Duffing} oscillators}.
\newblock \emph{\bibinfo{journal}{Chaos: An Interdisciplinary Journal of Nonlinear Science}} \textbf{\bibinfo{volume}{27}} (\bibinfo{year}{2017}).

\bibitem{deshaka2025realization}
\bibinfo{author}{Deshaka, S.}, \bibinfo{author}{Arun, R.}, \bibinfo{author}{Sathish~Aravindh, M.}, \bibinfo{author}{Venkatesan, A.} \& \bibinfo{author}{Lakshmanan, M.}
\newblock \bibinfo{title}{Realization of multiple-input and single-output logic gates in nonlinear systems}.
\newblock \emph{\bibinfo{journal}{Phys. Rev. E}} \textbf{\bibinfo{volume}{112}}, \bibinfo{pages}{014219} (\bibinfo{year}{2025}).

\bibitem{mathew2020synthetic}
\bibinfo{author}{Mathew, J.~P.}, \bibinfo{author}{Pino, J.~d.} \& \bibinfo{author}{Verhagen, E.}
\newblock \bibinfo{title}{Synthetic gauge fields for phonon transport in a nano-optomechanical system}.
\newblock \emph{\bibinfo{journal}{Nat. Nanotechnol.}} \textbf{\bibinfo{volume}{15}}, \bibinfo{pages}{198--202} (\bibinfo{year}{2020}).

\bibitem{delpino2022}
\bibinfo{author}{Del~Pino, J.}, \bibinfo{author}{Slim, J.~J.} \& \bibinfo{author}{Verhagen, E.}
\newblock \bibinfo{title}{Non-hermitian chiral phononics through optomechanically induced squeezing}.
\newblock \emph{\bibinfo{journal}{Nature}} \textbf{\bibinfo{volume}{606}}, \bibinfo{pages}{82--87} (\bibinfo{year}{2022}).

\bibitem{Bagheri2013Photonic}
\bibinfo{author}{Bagheri, M.}, \bibinfo{author}{Poot, M.}, \bibinfo{author}{Fan, L.}, \bibinfo{author}{Marquardt, F.} \& \bibinfo{author}{Tang, H.~X.}
\newblock \bibinfo{title}{Photonic cavity synchronization of nanomechanical oscillators}.
\newblock \emph{\bibinfo{journal}{Phys. Rev. Lett.}} \textbf{\bibinfo{volume}{111}}, \bibinfo{pages}{213902} (\bibinfo{year}{2013}).

\bibitem{aspelmeyer2014cavity}
\bibinfo{author}{Aspelmeyer, M.}, \bibinfo{author}{Kippenberg, T.~J.} \& \bibinfo{author}{Marquardt, F.}
\newblock \bibinfo{title}{Cavity optomechanics}.
\newblock \emph{\bibinfo{journal}{Rev. Mod. Phys.}} \textbf{\bibinfo{volume}{86}}, \bibinfo{pages}{1391--1452} (\bibinfo{year}{2014}).

\bibitem{Slim2026strong}
\bibinfo{author}{Slim, J.~J.} \& \bibinfo{author}{Verhagen, E.}
\newblock \bibinfo{title}{Strong nanomechanical {Duffing} nonlinearity and interactions induced through cavity optomechanics} (\bibinfo{year}{2026}).

\bibitem{leijssen2017nonlinear}
\bibinfo{author}{Leijssen, R.}, \bibinfo{author}{La~Gala, G.~R.}, \bibinfo{author}{Freisem, L.}, \bibinfo{author}{Muhonen, J.~T.} \& \bibinfo{author}{Verhagen, E.}
\newblock \bibinfo{title}{Nonlinear cavity optomechanics with nanomechanical thermal fluctuations}.
\newblock \emph{\bibinfo{journal}{Nat. Commun.}} \textbf{\bibinfo{volume}{8}}, \bibinfo{pages}{ncomms16024} (\bibinfo{year}{2017}).

\bibitem{bachtold2022mesoscopic}
\bibinfo{author}{Bachtold, A.}, \bibinfo{author}{Moser, J.} \& \bibinfo{author}{Dykman, M.}
\newblock \bibinfo{title}{Mesoscopic physics of nanomechanical systems}.
\newblock \emph{\bibinfo{journal}{Rev. Mod. Phys.}} \textbf{\bibinfo{volume}{94}}, \bibinfo{pages}{045005} (\bibinfo{year}{2022}).

\bibitem{dally2012digital}
\bibinfo{author}{Dally, W.~J.} \& \bibinfo{author}{Harting, R.~C.}
\newblock \emph{\bibinfo{title}{Digital design: a systems approach}} (\bibinfo{publisher}{Cambridge University Press}, \bibinfo{year}{2012}).

\bibitem{miller2009device}
\bibinfo{author}{Miller, D.~A.}
\newblock \bibinfo{title}{Device requirements for optical interconnects to silicon chips}.
\newblock \emph{\bibinfo{journal}{Proceedings of the IEEE}} \textbf{\bibinfo{volume}{97}}, \bibinfo{pages}{1166--1185} (\bibinfo{year}{2009}).

\bibitem{fong2012frequency}
\bibinfo{author}{Fong, K.~Y.}, \bibinfo{author}{Pernice, W. H.~P.} \& \bibinfo{author}{Tang, H.~X.}
\newblock \bibinfo{title}{Frequency and phase noise of ultrahigh $q$ silicon nitride nanomechanical resonators}.
\newblock \emph{\bibinfo{journal}{Phys. Rev. B}} \textbf{\bibinfo{volume}{85}}, \bibinfo{pages}{161410} (\bibinfo{year}{2012}).

\end{thebibliography}
\end{document}